\documentclass[usenatbib]{mn2e}

\usepackage{graphicx}
\usepackage{amsmath}
\usepackage{amssymb}
\usepackage{url}
\usepackage{multirow}
\usepackage{ctable}
\usepackage{hyperref}
\usepackage{wasysym}
\usepackage[T1]{fontenc}
\usepackage{aecompl}

\usepackage[compatibility=false]{caption}

\usepackage[rounding]{rccol}
\rcDecimalSign{.}

\def\apj{ApJ}
\def\apjl{ApJ}
\def\apjs{ApJS}
\def\aap{A\&A}
\def\mnras{MNRAS}
\def\nat{Nature}

\title[There is a short GRB in each long one]{There is a short
  gamma--ray burst prompt phase at the beginning of each long one}

\author[G. Calderone et al.]{G. Calderone$^{1}$\thanks{E-mail: {\tt giorgio.calderone@gmail.com}},
G. Ghirlanda$^{1}$,
G. Ghisellini$^{1}$,
M. G. Bernardini$^{1}$,
S. Campana$^{1}$,\newauthor
S. Covino$^{1}$,
D'Avanzo$^{1}$,
V. D'Elia$^{2,3}$,
A. Melandri$^{1}$,
R. Salvaterra$^{4}$,
B. Sbarufatti$^{5,1}$,\newauthor
G. Tagliaferri$^{1}$\\
$^{1}$INAF--Osservatorio Astronomico di Brera, via E. Bianchi 46, I--23807 Merate (LC), Italy\\
$^{2}$ASI--Science Data Centre, Via del Politecnico snc, I--00133 Rome, Italy\\
$^{3}$INAF--Osservatorio Astronomico di Roma, via Frascati 33, I--00040 Monteporzio Catone (RM), Italy\\
$^{4}$INAF--IASF Milano, via E. Bassini 15, I--20133, Milano, Italy\\
$^{5}$Department of Astronomy and Astrophysics, Pennsylvania State University, University Park, PA, 16802, USA\\
}

\begin{document}

\pagerange{\pageref{firstpage}--\pageref{lastpage}} \pubyear{2014}

\maketitle
\label{firstpage}

\begin{abstract}
  We compare the prompt intrinsic spectral properties of a sample of
  short Gamma--ray Bursts (GRBs) with the first 0.3 seconds (rest
  frame) of long GRBs observed by {\it Fermi}/GBM.  We find that short
  GRBs and the first part of long GRBs lie on the same $E_{\rm
    p}$--$E_{\rm iso}$ correlation, that is parallel to the relation
  for the time averaged spectra of long GRBs.  Moreover, they are
  indistinguishable in the $E_{\rm p}$--$L_{\rm iso}$ plane. This
  suggests that the emission mechanism is the same for short and for
  the beginning of long events, and both short and long GRBs are very
  similar phenomena, occurring on different timescales.  If the
  central engine of a long GRB would stop after $\sim 0.3 \times
  (1+z)$ seconds the resulting event would be spectrally
  indistinguishable from a short GRB.
\end{abstract}


\section{Introduction}
\label{sec:intro}

Gamma--ray bursts (GRBs) are transient emission episodes of radiation
detected at high energies.  The first emission phase, detected at hard
X--rays and $\gamma$--rays, lasts for $\sim$~0.01 ms--100 s (prompt
phase).  Then, the bulk of emitted radiation shifts to lower energies
and becomes observable at longer wavelengths, from X--rays to radio,
with typical duration of $\sim$~days--months (afterglow phase).  The
observed duration of the prompt phase is characterised by the $T_{90}$
parameter, i.e. the time interval during which the central 90\% of the
counts are recorded by the detector.  The distribution of $T_{90}$ of
GRBs observed by the Burst And Transient Source Experiment (BATSE) on
board the Compton Gamma Ray Observatory ({\it GCRO}) has been found to
be bimodal with a separation at $\sim$~2 s in the observer frame
\citep{1993-Kouvelioutou-shortVsLong}.  According to this finding,
GRBs are classified either as {\it short} gamma--ray burst (SGRB) if
$T_{90} <$~2, or as {\it long} ones (LGRB) if $T_{90} >$~2 s (but see
\citealt{2013-Bromberg-ShortVsLongCollapsars}).  Besides, the prompt
phase of SGRBs is characterised by harder spectra
\citep{1993-Kouvelioutou-shortVsLong} and smaller spectral lags
between different energy bands \citep{2000-Norris-SpecLag} with
respect to the prompt phase of LGRBs.

For bursts with reliable redshift estimates, it has been shown that
SGRBs are systematically less energetic than LGRBs, with total X and
$\gamma$--ray emitted energies smaller by a factor $\sim$~10--100
\citep{2009-Ghirlanda-ShortVsLong}.  Also, the afterglows of SGRBs,
when detected, are correspondingly dimmer than those of LGRBs, but
similar in other respects \citep{2008-Gehrels-PromptAfterglow,
  2013-Margutti-PromptAfterglowConn, 2014-DAvanzo-SBAT8}.  Finally,
several nearby ($z<0.5$) long GRBs have been associated with
explosions of core--collapse supernovae
\citep{2012-Hjorth-GRB-SN-Connection}, while there is no similar
evidence for short bursts \citep{2013-Berger-SGRB}.  These findings
suggest that short and long GRBs might originate from different
progenitors (\citealt{2006-Meszaros-GRB, 2013-Berger-SGRB}).

Observationally, the most important difference between short and long
GRBs is their $T_{90}$ duration.  A first attempt to compare the
spectral properties of short and long GRBs detected by {\it
  CGRO}/BATSE showed that (i) the difference in hardness could be due
to a harder low energy spectral index of short GRBs rather than a
harder peak energy and (ii) that the spectra of SGRBs and the first
1--2 s of LGRBs appear similar \citep{2004-Ghirlanda-SpectraSGRB}.
These results suggested that the engine might be similar in the two
classes, but the activity would last longer in the case of LGRBs
\citep{2010ApJ...725..225G}.  Also, \citet{2002-Nakar-TempoPropSGRB}
found that the ratio of the shortest pulse duration to the total burst
duration for both short and the first 1--2 s of long GRBs were
comparable.

With the advent of the Gamma Burst Monitor (GBM) on board {\it Fermi},
it became possible to compare the spectral properties of large samples
of short and long GRBs and to compare them with those detected by {\it
  CGRO}/BATSE.  \citet{2011-Nava-GBMBatseComparison} showed that long
and short GRBs occupy different regions in the observer frame hardness
(defined by the peak of the $\nu F_\nu$ spectrum) versus fluence, with
SGRBs having smaller fluences than long events. This also suggested
that the possible selection of fluence limited samples for the
comparison of SGRBs and LGRBs could introduce biases.

The availability of redshift estimates for long GRBs allowed one to
estimate their rest frame (intrinsic) spectral properties, and to
highlight a few correlations among them (see
\citealt{2006-GhirlandaRev-GRBCorrCosmo} for a review).
\citet{2002-Amati-rel} found that the rest frame $\nu L_{\nu}$ peak
energy ($E_{\rm p}$) is correlated with the total energy emitted in
the 1 keV--10 MeV energy range (under the hypothesis of isotropic
emission, $E_{\rm iso}$), with a slope of $\sim$~0.5.
\citet{2004-Yonetoku-EpLiso} found a correlation between $E_{\rm p}$
and the isotropic peak luminosity evaluated at the flux peak over an
interval of 1 s ($L_{\rm p,iso}$), with a slope of $\sim$~0.4.  The
latter correlation is valid also when considering the time resolved
spectral quantities $E_{\rm p}(t)$ and $L_{\rm iso} (t)$ of a single
burst, i.e. the evolutionary tracks of GRB spectra in the $E_{\rm
  p}$--$L_{\rm iso}$ plane align with the Yonetoku relation
\citep{2009-Firmani-TimeResolvedCorrel,
  2010-Ghirlanda-TimeResolvedFermi, 2012-Frontera-TimeResolvCorr}.

With the fast slewing {\it Swift} satellite \citep{2004-Gehrels-Swift}
it became possible to localize the X--ray afterglows of short GRBs,
and estimate their redshifts by means of the associated host galaxies
\citep{2005-Gehrels-ShortAfterglow}.  The comparison of intrinsic
spectral properties of short and long GRBs have shown that short GRBs
are consistent with the Yonetoku relation, but are significant
outliers of the Amati relation \citep{2006-Amati-rel-update,
  2008-Amati-rel-5SGRB, 2009-Ghirlanda-ShortVsLong,
  2014-DAvanzo-SBAT8}.  However, by analyzing a sample of 7 short
GRBs, \citet{2012-ZhangRevisitingLongShortClassif} suggest that short
GRBs might follow a parallel Amati relation at lower values of $E_{\rm
  iso}$.  Moreover, the SGRBs follow the same three parameter
correlation ($E_{\rm X,iso}$--$E_{\rm \gamma,iso}$--$E_{\rm p}$) valid
for long GRBs \citep{2012-Bernardini-UnivScalingRel3Param,
  2013-Margutti-PromptAfterglowConn}.  The isotropic luminosities are
similar in both short and long GRBs, but the former are less energetic
than the latter by a factor similar to the ratio of their durations.
When considering the time averaged spectra, short GRBs have harder
low--energy spectral index, but this difference vanishes when
comparing the SGRBs with only the first 1--2 s of long GRBs
\citep{2009-Ghirlanda-ShortVsLong}.

Also the time resolved spectroscopy has shown that the observed peak
energy tracks the flux evolution in both short and long GRBs
\citep{2010ApJ...725..225G, 2011-Ghirlanda-ShortVsLong}, suggesting a
common physical mechanism linking these quantities.  The existence of
a time resolved correlation between $E_{\rm p}(t)$ and $L_{\rm
  iso}(t)$ was also shown to hold in short GRBs
\citep{2011-Ghirlanda-ShortVsLong}.  This is the most compelling
evidence that the $E_{\rm p}(t)$--$L_{\rm iso}(t)$ correlation holding
in long and short GRBs (with similar slope and normalization) hints to
a common origin which could be related to the emission mechanism
\citep{2011-Ghirlanda-ShortVsLong} and that the corresponding Yonetoku
correlation (holding between time integrated properties) cannot be
subject to strong selection effects.  An interesting hypothesis
discussed in \citet{2009-Ghirlanda-ShortVsLong,
  2011-Ghirlanda-ShortVsLong, 2013ApJ...770...32G} is that both short
and long GRBs may share a common emission process, and that the
observed differences may be ascribed to the different engine lifetime
of their progenitors.

Yet, the comparison of short and long GRBs in search for possible
similarities or differences should account for their possible
different redshift distributions.  While several LGRBs have their
redshift measured, the population of short bursts still suffers from a
lack of redshift measures. However, recent collection of small, well
defined, samples of SGRBs with measured redshifts
\citep[e.g.][]{2014-DAvanzo-SBAT8} allowed us to compare the energetic
properties of short and long events in their rest frame.

The aim of this work is to further explore the similarities between
short and long GRBs by comparing their intrinsic (i.e. rest frame)
spectral properties estimated on the same rest frame time scales.  The
average $T_{90} / (1+z)$ duration of the short GRBs with reliable
(spectroscopic) redshifts and without X--ray extended emission in the
\citet{2014-DAvanzo-SBAT8} sample is 0.3 s (10 bursts).  This will be
our reference time scale to perform spectral analysis of the first
part of long GRBs, and compare the results with those of short GRBs.

Throughout the paper, we assume a $\Lambda$CDM cosmology with H$_0$ =
71 km s$^{-1}$ Mpc$^{-1}$, $\Omega_{\rm m}$ = 0.27, $\Omega_\Lambda$ =
0.73.

\section{The sample}
\label{sec-sample}

Since we aim to study the prompt emission spectral properties and
energetic/luminosity of GRBs, we need a broad energy coverage in order
to determine where the peak energy is. While {\it Swift}/BAT has a
limited energy range (15--150keV) which is not suited for GRB prompt
emission spectral characterization, the GBM instrument on board {\it
  Fermi} covers almost 2 orders of magnitude in energy with the NaI
detectors (8keV--1MeV) and can extend this energy range to a few tens
of MeV with the inclusion of the data of the BGO detectors.  Hence we
selected all GRBs observed by {\it Fermi}/GBM up to December 2013 with
a redshift estimate.  This amounts to 64 long and 7 short GRBs.

Among the long ones we discarded:
2 GRBs with missing response matrix files; 
2 GRBs observed with a non--standard Low Level Threshold\footnote{\href{http://fermi.gsfc.nasa.gov/ssc/data/access/gbm/llt\_settings.html}{http://fermi.gsfc.nasa.gov/ssc/data/access/gbm/llt\_settings.html}};
3 GRBs whose first part was missed by the GBM;
12 GRBs for which we could not constrain either the low energy
spectral index or the peak energy (\S\ref{sec-data_an}).  The final
LGRB sample comprises 45 long bursts.

{\it Fermi}/GBM observed 7 short GRBs with known redshift.  To this
sample we added the SGRB flux limited sample of 12 sources with
redshift discussed in \citet[][hereafter D14
  sample]{2014-DAvanzo-SBAT8}, but discarded: GRB 080905A since its
redshift is likely not accurate, GRB 090426 and GRB 100816A since
their classification as short GRB is debated.  Four GRBs in the D14
sample were also in the GBM sample: for these bursts we considered the
results reported in D14.  The final SGRB sample comprises 3 GRB
observed with {\it Fermi}/GBM and 9 from D14.

The short GRB sample, although relatively small, stems from a flux
limited sample of short GRB with a redshift completeness of $\sim$70\%
\citep{2014-DAvanzo-SBAT8}.  We dropped three burst from this sample,
hence the redshift completeness drops to $\sim$~60\%, but we added
three more bursts detected by {\it Fermi}/GBM.  The distributions of
low--energy spectral index in both our short sample and the
corresponding one\footnote{Bursts with $T_{90} < 2$ s and a either a
  Band or cutoff power law best fitting model in the
  \citet{2014-Gruber-GBMCat-4yr} catalog: 70 GRB.} in the
\citet{2014-Gruber-GBMCat-4yr} catalog are actually indistinguishable
(K--S test probability: 0.47).  Since the spectral index is a redshift
independent property we assume that our short GRB sample is a
reasonably good representation of the parent distribution of short
GRBs observable with currently available detectors.

\begin{figure}
  \includegraphics[width=.45\textwidth]{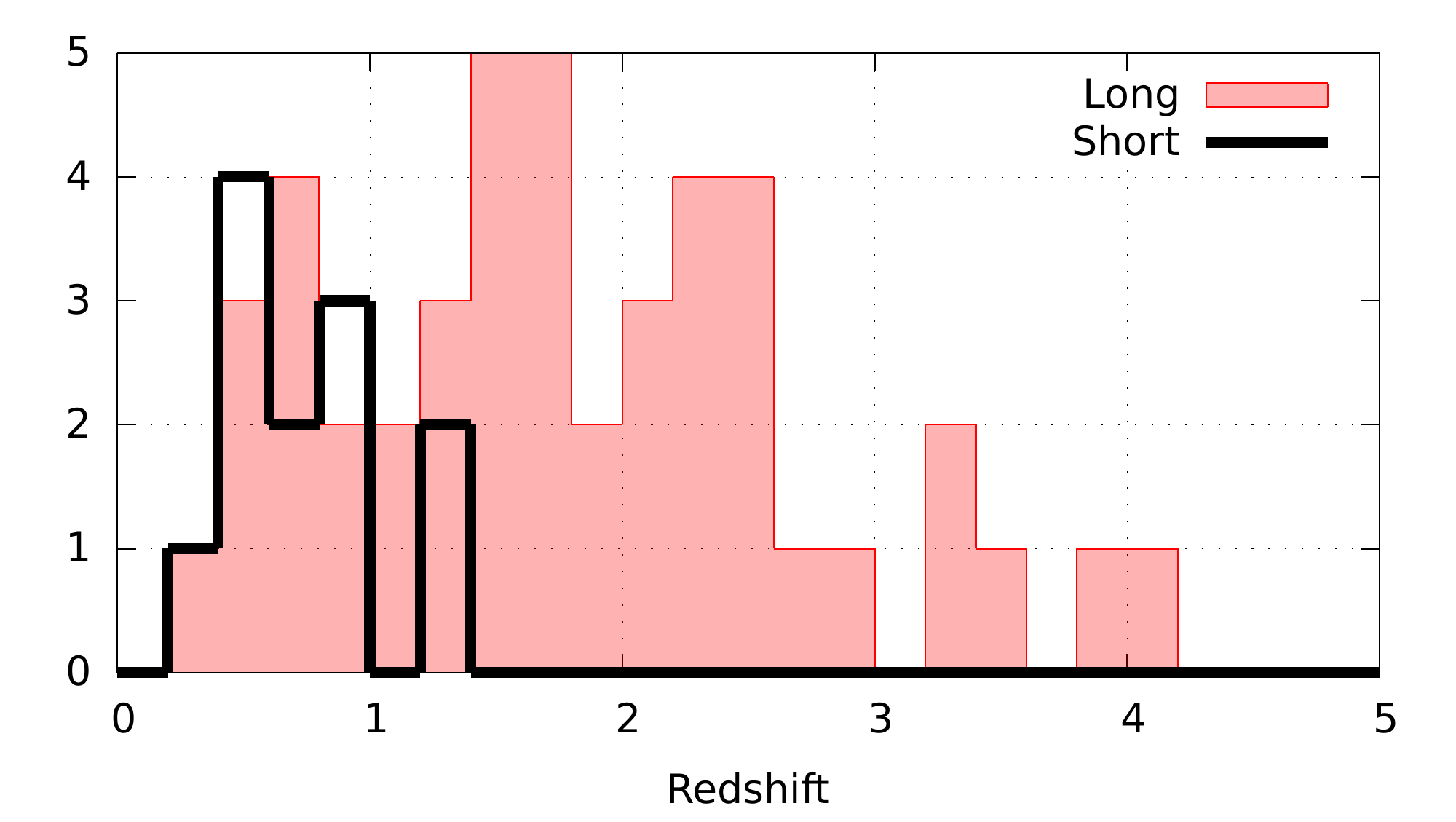}
  \caption{Redshift distribution for both the SGRB (12) and LGRB (45)
    bursts (references for redshift estimates are given in
    Tab.\ref{tab1}).}
  \label{fig-z}
\end{figure}

The total (LGRB+SGRB) comprises 57 bursts (Tab.\ref{tab1}).  
Fig. \ref{fig-z} shows the redshift distribution for both the SGRB and
LGRB samples (references for redshift estimates are given in
Tab.\ref{tab1}).

\section{Data analysis}
\label{sec-data_an}

Our spectral analysis aims at estimating the intrinsic peak energy
($E_{\rm p}$), isotropic equivalent luminosity ($L_{\rm iso}$) and
emitted energy ($E_{\rm iso}$) for the GRBs in our sample.  For the 3
short GRBs observed by {\it Fermi}/GBM we performed a spectral
analysis on the entire duration of the burst ({\em short} analysis).
For the remaining 9 SGRB we considered the spectral properties
relative to the time integrated emission reported in the D14 paper
({\it short D14}).  For the 45 long GRBs we perform two different
spectral analysis: one for the first 0.3 s in the rest frame
(corresponding to $0.3 \times (1+z)$ s in the observer frame, {\em
  first} analysis) and one for the whole duration of the burst ({\em
  whole} analysis).  All $E_{\rm iso}$ and $L_{\rm iso}$ quantities
are evaluated in the (rest frame) energy range 1 keV--10 MeV.

All data analysis has been carried according to the
procedure outlined below.

\subsection{Detectors, energy selection and background fitting}

For each GRB we selected the most illuminated NaI detector(s), and the
corresponding BGO one.  The BGO detector is always included, even if
there is no significant detection above background.  The energy
selection is in the range 8--800 keV for NaI detectors, and 200
keV--35 MeV for BGO ones.  Systematic residuals at $\sim$~33 keV of
the NaI
detectors\footnote{\href{http://fermi.gsfc.nasa.gov/ssc/data/analysis/GBM\_caveats.html}{http://fermi.gsfc.nasa.gov/ssc/data/analysis/GBM\_caveats.html}}
are neglected.

For each channel of all detectors we perform a polynomial fit (up to
the third order) to the observed background count rate in the CSPEC
files,\footnote{Time binned count spectra with time resolution of
  1.024 s from the burst trigger time $T_0$ to $T_0$+600 s, and time
  resolution of 4.096 s for a few thousands seconds before and after
  the burst.} on a time interval before and after the burst longer
than the burst duration (typically $\gtrsim$~100 s).  The length of
the background time intervals is progressively increased until the
uncertainties on the expected background counts during the burst
becomes smaller than their intrinsic statistical fluctuations.  This
approach provides an objective way to select the background time
intervals.  We also checked that the background fit provide an
adequate fit for all energy channels by means of $\chi^2$
goodness--of--fit test.  For long bursts, we used exactly the same
background model for both the {\it first} and {\it whole} analysis.

The detectors used and the background time selections for each burst
are shown in Tab. \ref{tab1}.

\subsection{Time selection}

For the GRB spectral analysis we used the TTE data files\footnote{The
  list of all recorded counts with time and channel tags (Time Tagged
  Events).  Data are available from $\sim T_0 - 25$ s to $\sim T_0 +
  300$ s.} to select the counts in the appropriate time intervals:
either the first 0.3 seconds (rest frame) for the {\em first}
analysis, or the whole burst duration for both the {\em short} and
{\em whole} analysis.

For the {\it short} and {\it whole} analysis of GRBs present in both
our sample and the \citet{2014-Gruber-GBMCat-4yr} catalog we consider
their time selection.  This choice allows us to compare our results
with those of \citet{2014-Gruber-GBMCat-4yr}, as discussed in
\S\ref{cmp-g14}.  For the other bursts the time selection was
performed by a visual inspection of the count rate light curves.

For the {\em first} analysis we searched for the first occurrence of a
0.3 s long (rest frame) time bin in which the counts in all NaI
detectors were significantly (at 3$\sigma$ level) above the expected
background.  The search has been performed with a 0.2 s resolution
starting at 10 s before the trigger time.

The time selections for each burst are shown in Tab. \ref{tab2}.

\subsection{Spectral fitting}
\label{sec-specfit}

The GRB spectral models used for spectral analysis are a modified
version\footnote{The \texttt{XSPEC} implementation of this spectral
  model is available
  \href{http://www.giorgiocalderone.url.ph/xspec\_ggrb.tar.gz}{http://www.giorgiocalderone.url.ph/xspec\_ggrb.tar.gz}.}
of either the cutoff power law or the Band model \citep{1993-Band}, in
which the free parameters are:
\begin{itemize}
\item \texttt{log\_Ep}: the logarithm of the $\nu F_{\nu}$ peak energy
  in keV;
\item \texttt{alpha}: the photon spectral index for energies smaller
  than the peak energy;
\item \texttt{beta} (only for the Band model): the photon spectral
  index for energies greater than the peak energy;
\item \texttt{log\_F}: the logarithm of the integrated flux in the
  rest frame energy range 1 keV--10 MeV.
\end{itemize}
The spectral indices are bounded to be \texttt{alpha} $>-2$ and $-6 <
$ \texttt{beta} $ <-1.7$.  For the {\em whole} spectral analysis we
also included detectors effective area correction as free parameters,
bounded in the range 0.5--2.  Whenever the resulting area corrections
are not constrained we set all calibration factors to one.  The
\texttt{log\_F} parameter is used to estimate the intrinsic isotropic
luminosity $L_{\rm iso} = 4 \pi D_{\rm L}^2 \times F$, without the
need to propagate the uncertainties on the other parameters.  Finally,
isotropic emitted energy is estimated as $E_{\rm iso} = L_{\rm iso}
\Delta T_{\rm rest}$, where $\Delta T_{\rm rest} \times (1+z)$ is the
spectrum integration time.

The spectral model is folded with the detector response matrix, summed
with the background counts expected in the same time interval, and
compared to the observed counts by means of the Cash statistic
\citep{1979-Cash-Cstat} with Castor normalization (C--STAT).  The
model fitting is performed with \texttt{xspec} ver. 12.8.1g
\citep{1996-Arnaud-xspec} by minimizing the C--STAT value.  We always
used the detector maximum energy resolution, i.e. we did not rebin the
channels.

The choice of the spectral model (cutoff power law or Band) is
performed according to the following criterion: for each burst we
started with the Band model with both spectral indices free to vary in
the minimization process.  If the \texttt{beta} parameter uncertainty
is larger than a nominal threshold of 0.5, but still significantly
lower than the \texttt{alpha} parameter, we fixed \texttt{beta} to its
typical value, namely --2.3 \citep{1993-Band,
  2002-Ghirlanda-TimeResolvSpecBrightGRB}, and repeat the fit.  If
\texttt{beta} hits the lower limit (--6) we use the cutoff power law
model instead of the Band model.  If \texttt{beta} is $>-2$ and
\texttt{alpha}$>$\texttt{beta} we consider the resulting $E_{\rm p}$
and $L_{\rm iso}$ as lower limits.  The true location of the $\nu
F{\nu}$ peak likely lies on the extrapolation of the spectrum actually
constrained by the data.  By assuming \texttt{alpha}=--1 \citep[the
  typical value for this parameter, ][]{2011-Nava-438FermiGRB}, this
extrapolation lies on a line of slope 1 in the $E_{\rm p}$--$L_{\rm
  iso}$ plane.

In 12 cases we could not detect a curvature in the spectrum,
i.e. we could not constrain either the \texttt{alpha} or the
\texttt{log\_Ep} parameters.  These bursts were discarded from our
sample (\S\ref{sec-sample}).

The parameter uncertainties (quoted at 1$\sigma$) for the {\em whole}
analysis are estimated with the usual $\Delta \chi^2$ method
\citep{1976-Avni-DeltaChisqUncert, 1976-Cash-DeltaChisqUncert}.  For
the {\em short} and {\em first} analysis we adopted a different
approach since the counts in the high--energy channels of the
detectors are often very low.  In these cases we start by performing a
fit in the usual way, and use the best fit parameter estimates to
simulate several data sets for each detector (using the
\texttt{fakeit} command).  Then we run the fitting process on the mock
data sets, and consider the distribution of the resulting best fit
parameters.  The final uncertainties are estimated as the central
interval containing 68.3\% of the best fit values.  The simulation
iterates until the lower and upper limits of the confidence interval
change by less than 5\%.  Typically 400--600 simulations are required
to satisfy the convergence criterion.  This Monte Carlo method is
described in detail in \citet[][their Sect. 15.6.1]{2007-nr-3rd}.

In \S\ref{cmp-g14} we compare the results of our {\it whole} analysis
to those of \citet{2014-Gruber-GBMCat-4yr}, for the bursts present in
both samples, and show that the two methods produce very similar
results.  However, our method ensures a homogeneous approach in all
our spectral analysis: we established an objective criterion to select
the background time intervals, and used exactly the same background
model in both the {\em first} and {\em whole} analysis.  The use of
logarithmic quantities in our spectral model results in simpler and
more symmetric parameter uncertainties, with respect to their linear
counterparts \citep[e.g.][]{2007-Cabrera-SpAnalysisSwift-wZ}.  Also,
the use of the integrated flux as model parameter, instead of the flux
at a given energy, allows us to directly evaluate the uncertainties on
$L_{\rm iso}$, avoiding the necessity to estimate the parameter
covariance matrix for error propagation.  Finally, the use of Monte
Carlo simulations in the {\em short} and {\em first} analysis provide
reliable parameter uncertainties even in the low count regime, when
the assumption that the C--STAT value is drawn from a $\chi^2$
distribution is not reliable.

\subsection{Results}

\begin{figure*}
  \includegraphics[width=.48\textwidth]{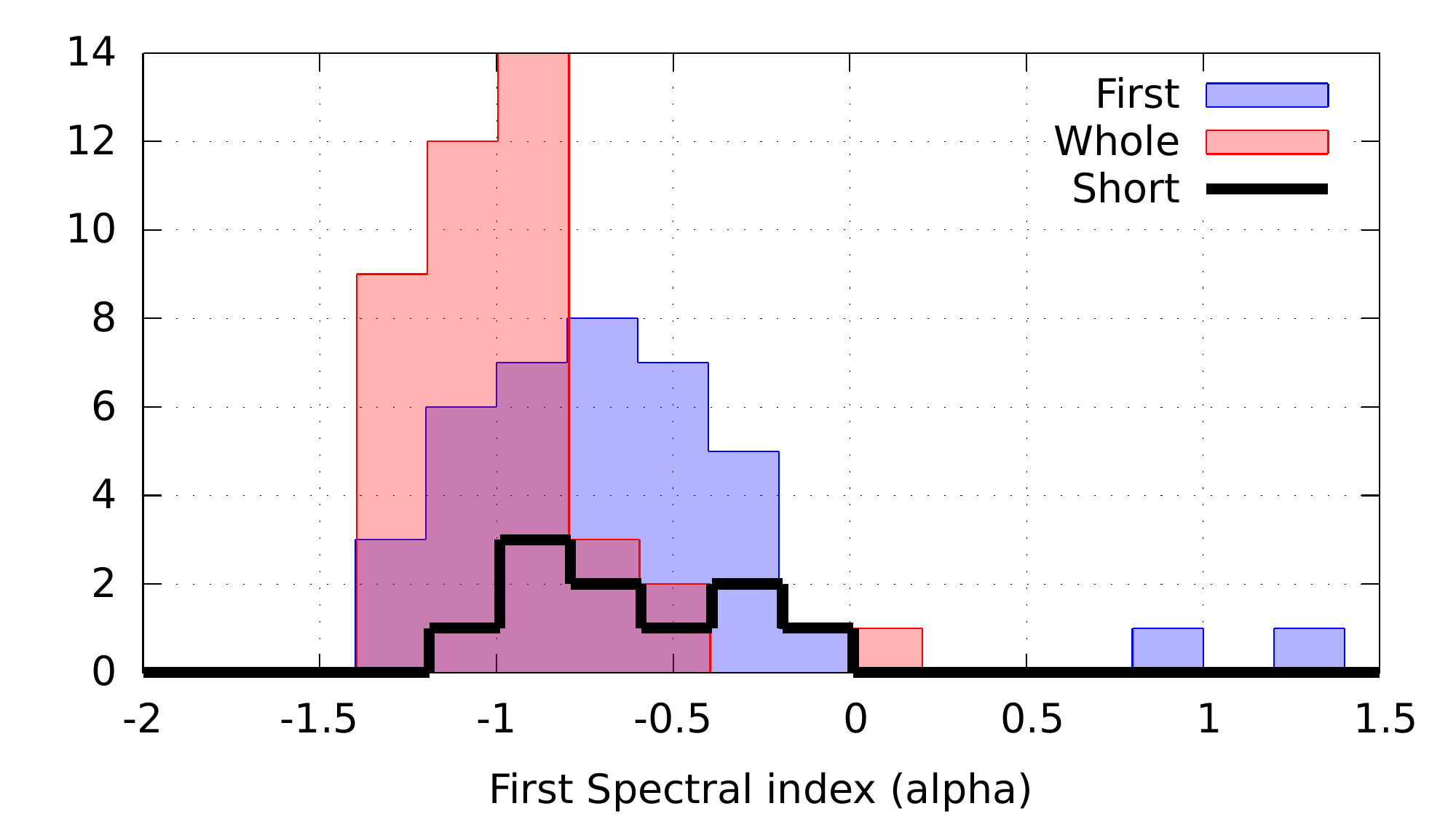}
  \includegraphics[width=.48\textwidth]{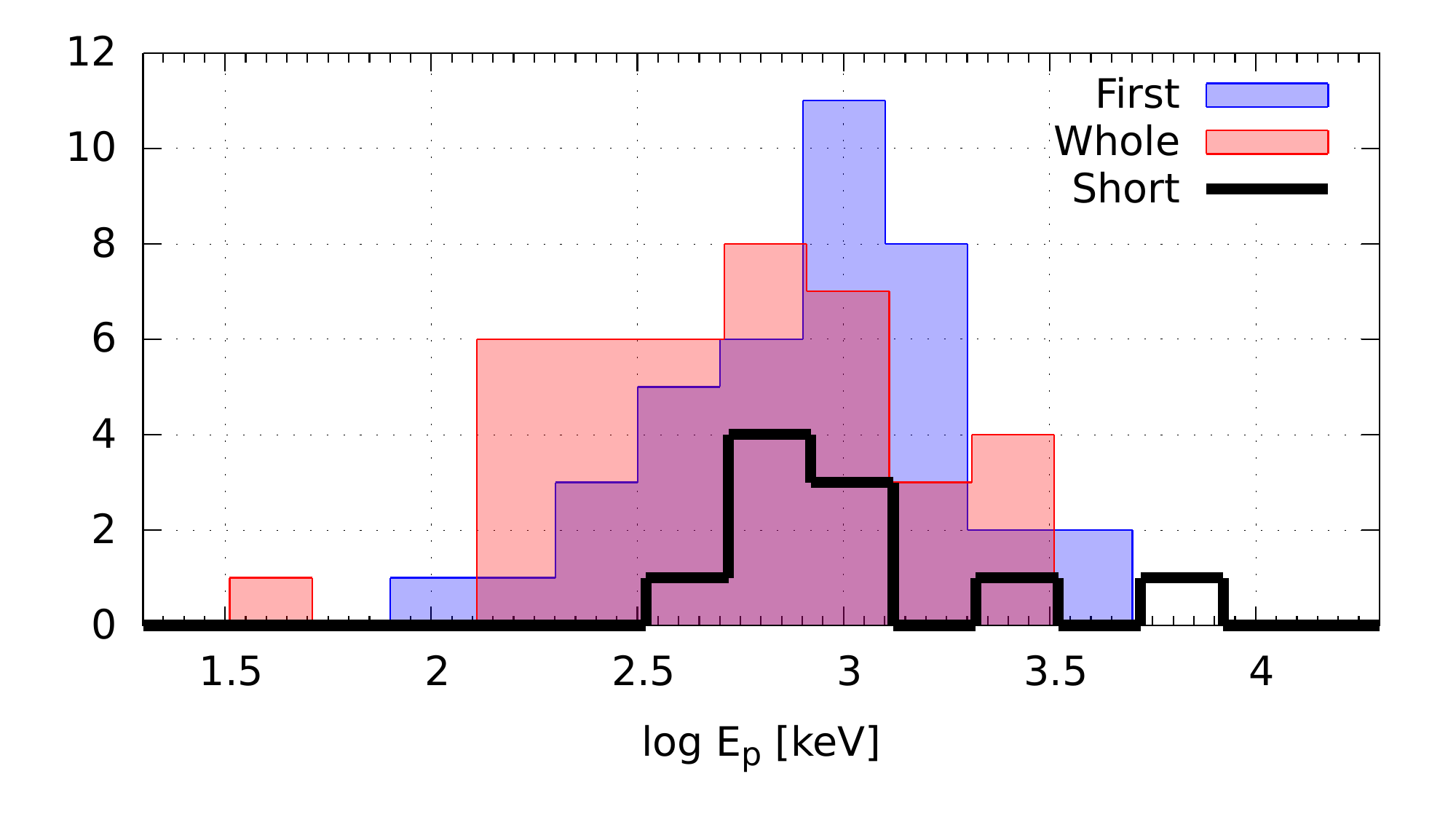}\\
  \includegraphics[width=.48\textwidth]{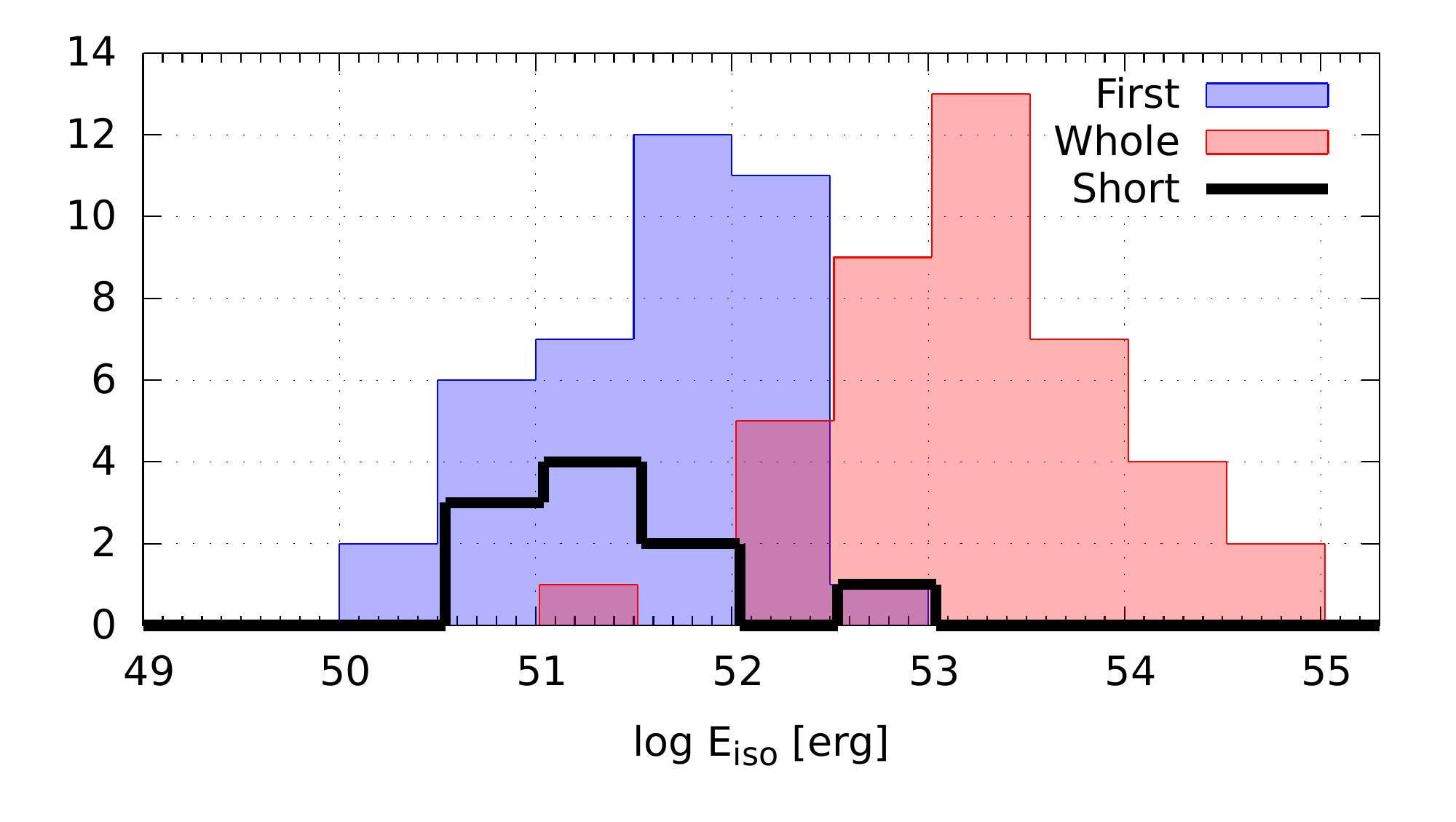}
  \includegraphics[width=.48\textwidth]{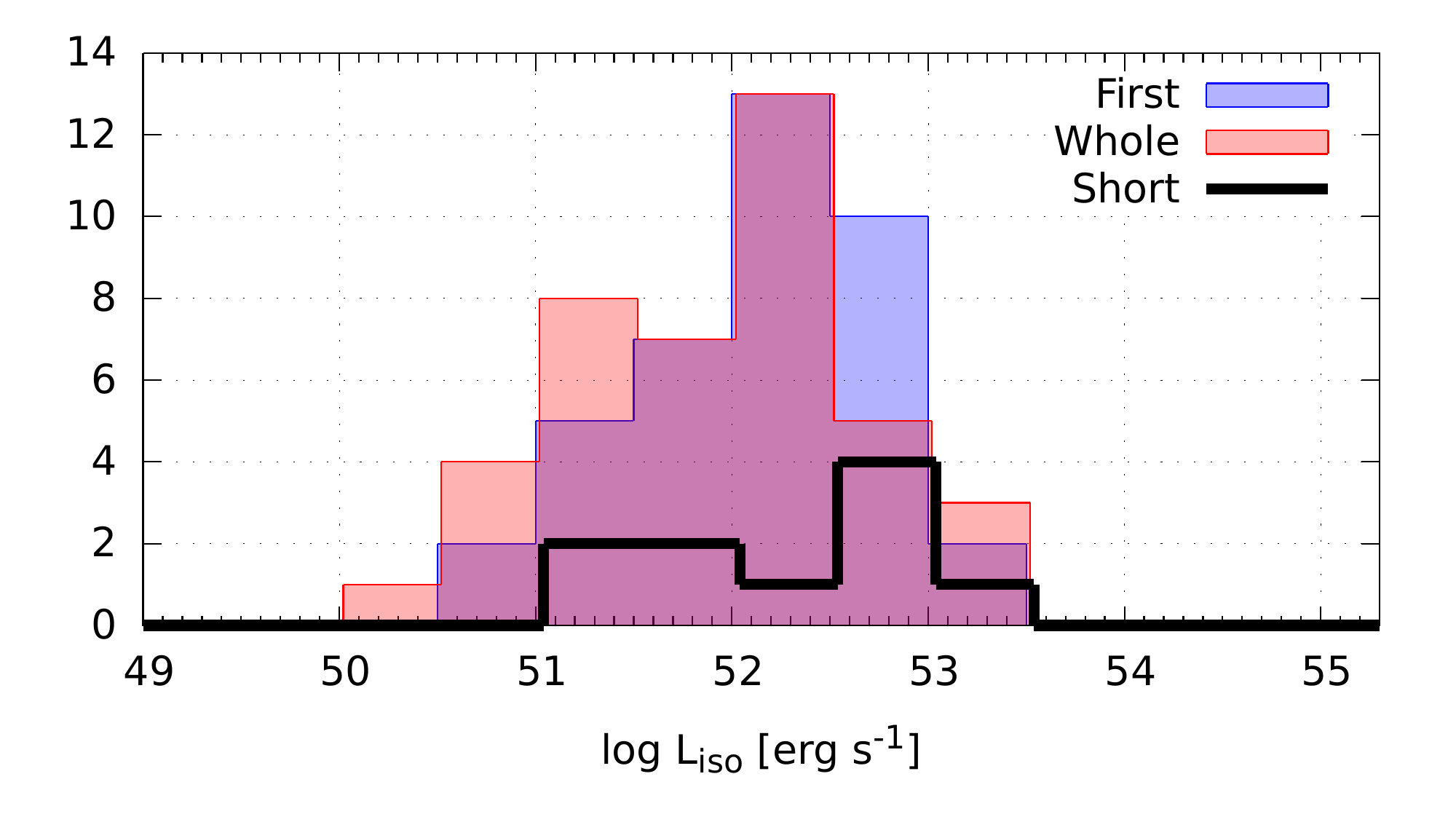}
  \caption{Histograms of relevant results of the spectral analysis.
    Upper panels: the low energy spectral index \texttt{alpha} (left);
    intrinsic $\nu L_{\nu}$ peak energy $E_{\rm p}$ (right).  Lower
    panels: isotropic equivalent, emitted energy $E_{\rm iso}$ (left)
    and luminosity $L_{\rm iso}$ (right), integrated in the 1 keV--10
    MeV energy range (rest frame).  The lower limits for $E_{\rm p}$
    and $E_{\rm iso}$ and the analysis on precursors are not accounted
    for in the histograms.}
  \label{fig-histo}
\end{figure*}

The results of spectral analysis, as well as the spectral quantities
reported in D14 for the short GRB sample, are shown in
Tab. \ref{tab2}.  The relevant quantities for the {\em short}, {\em
  first} and {\em whole} sub samples are shown in
Fig. \ref{fig-histo}.  The lower limits for $E_{\rm p}$ and $E_{\rm
  iso}$ (2 in the {\it short}, 4 in the {\em first} and 3 in the {\em
  whole} analysis, respectively) are not accounted for in the
histograms.

\begin{figure*}
  \includegraphics[width=.48\textwidth]{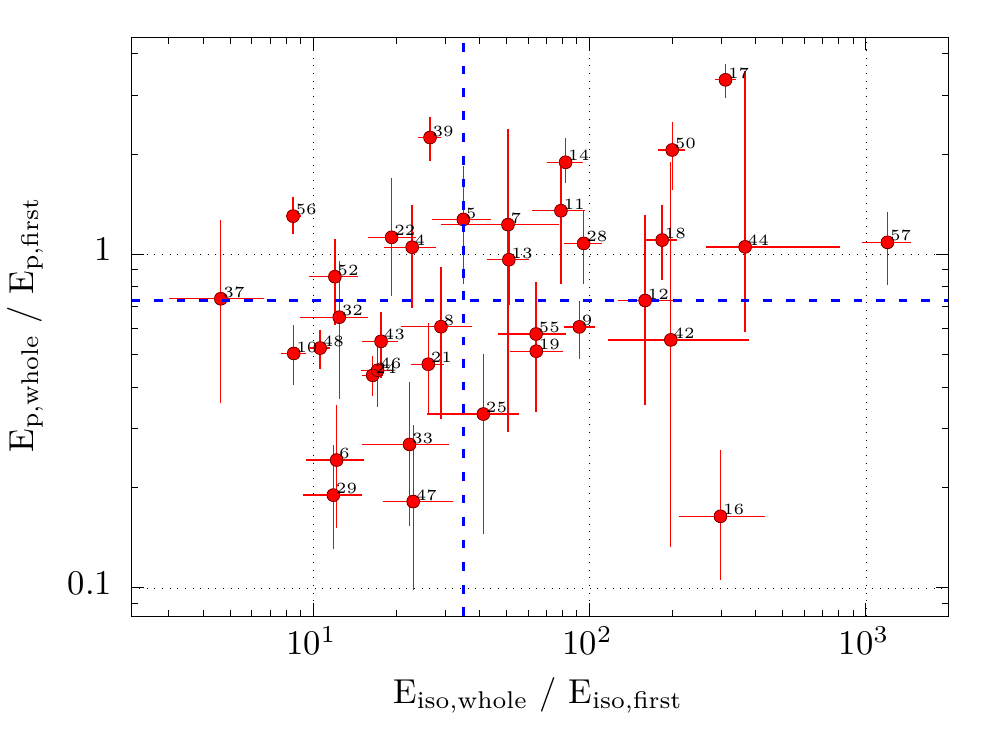}
  \includegraphics[width=.48\textwidth]{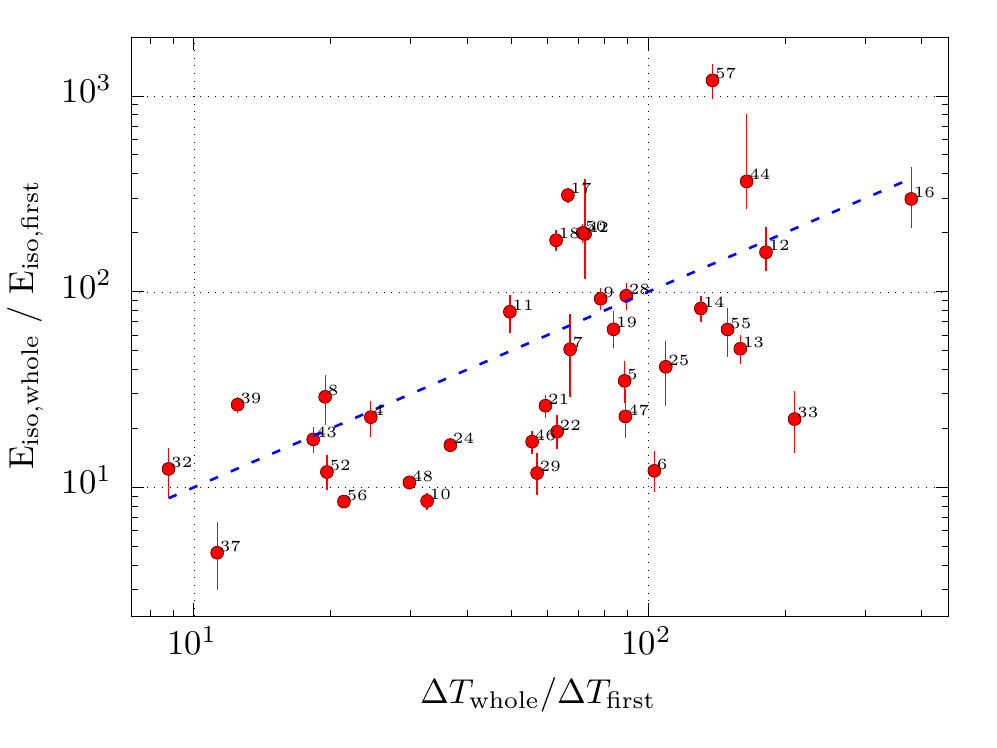}
  \caption{Left panel: ratio of {\em whole} to {\em first} peak energy
    vs. the same ratio for $E_{\rm iso}$.  The blue dashed lines are
    the median values of both ratios.  Right panel: the $E_{\rm iso,
      whole} / E_{\rm iso, first}$ ratio vs.  $\Delta T_{\rm whole} /
    \Delta T_{\rm first}$.  The blue dashed line is the 1:1 line.  The
    numbers beside the symbols are the GRB identifiers shown in Table
    \ref{tab1} and \ref{tab2}.}
  \label{fig-ratio}
\end{figure*}

Fig. \ref{fig-ratio} (left panel) shows the ratio of {\em first} to
{\em whole} peak energy vs. the same ratio of $E_{\rm iso}$.  The blue
dashed lines are the median values of both ratios.  The right panel
shows the $E_{\rm iso, whole} / E_{\rm iso, first}$ ratio vs.  $\Delta
T_{\rm whole} / \Delta T_{\rm first}$.  The blue dashed line is the
1:1 line.  The numbers beside the symbols are the GRB identifiers
shown in Table \ref{tab1} and \ref{tab2}.

Finally, in Fig. \ref{fig-corr1} and \ref{fig-corr2} we show the
location of all bursts in the $E_{\rm p}$--$E_{\rm iso}$ and $E_{\rm
  p}$--$L_{\rm iso}$ planes respectively.  The lower limits are shown
with arrows of slope 1, as discussed in \S\ref{sec-specfit}.

\subsection{Notes on individual bursts}
\label{sec-notes}
\begin{itemize}
\item GRB 091024 \citep{2011-Gruber-091024,
  2014-Nappo-Afterglow091024}: the GBM data are separated in two burst
  intervals, hence this GRB appear twice in Tab. \ref{tab1} (ID 23).
  The first interval is actually a precursor and it is analysed
  according to the {\em first} prescription.  The second interval
  comprises a second precursor and the main event.  The {\em first}
  analysis at the beginning of the main event did not provide reliable
  constraints on the peak energy, hence we consider only the {\em
    whole} analysis.

\item GRB 110213: for this burst the {\em first} analysis did not
  provide reliable constraints on the peak energy because the signal
  is significantly background dominated, hence we consider only the
  {\em whole} analysis.

\item GRB 120711A and GRB 120716A show a precursor in their light
  curve.  For these bursts we analysed the precursor spectra according
  to the {\em first} analysis.

\item GRB 130427A: the GBM data are unreliable after $\sim$~4 seconds
  from the trigger since the large amount of recorded events, due to
  the exceptional brightness of this burst, saturated the available
  bandwidth \citep{2014-Preece-GBM130427}.  Hence we consider only the
  {\em first} analysis for this burst.
\end{itemize}

By taking into account these notes the final subsamples comprises:
\begin{eqnarray}
  \nonumber
  Short: \ {\rm 12\ bursts,\ with\ 2\ lower\ limits}\\
  \nonumber
  First: \ {\rm 43\ bursts,\ with\ 4\ lower\ limits}\\
  \nonumber
  Whole: \ {\rm 44\ bursts,\ with\ 3\ lower\ limits}
\end{eqnarray}

\subsection{$E_{\rm p}$--$E_{\rm iso}$ and $E_{\rm p}$--$L_{\rm iso}$ correlations}

We use the results of the spectral analysis to test the
spectral--energy correlations in the $E_{\rm p}$--$E_{\rm iso}$ and
$E_{\rm p}$--$L_{\rm iso}$ planes.  The former is the Amati relation,
while the second is only similar to the Yonetoku relation, since we
use the $L_{\rm iso}$ values estimated on the time averaged spectra,
rather than the peak isotropic luminosity $L_{\rm p,iso}$
\citep{2004-Yonetoku-EpLiso}. 

We estimate the Spearman rank correlation coefficients and the
associated chance probability for the {\em short}, {\em first} and
{\em whole} results.  Also, we estimate the best fit correlations by
applying the unweighted bisector method
\citep{1990-Isobe-LinearRegression}.  Lower limits and precursor data
are not considered in this analysis.  The histograms of the residuals
from the best fit line, once projected on a scale perpendicular to the
line itself, are fitted with a Gaussian function in order to estimate
the scatter ($\sigma_{\rm sc}$) from the best fit.  Results are shown
in Tab.\ref{tab-corr}.

\begin{figure*}
  \includegraphics[width=\textwidth]{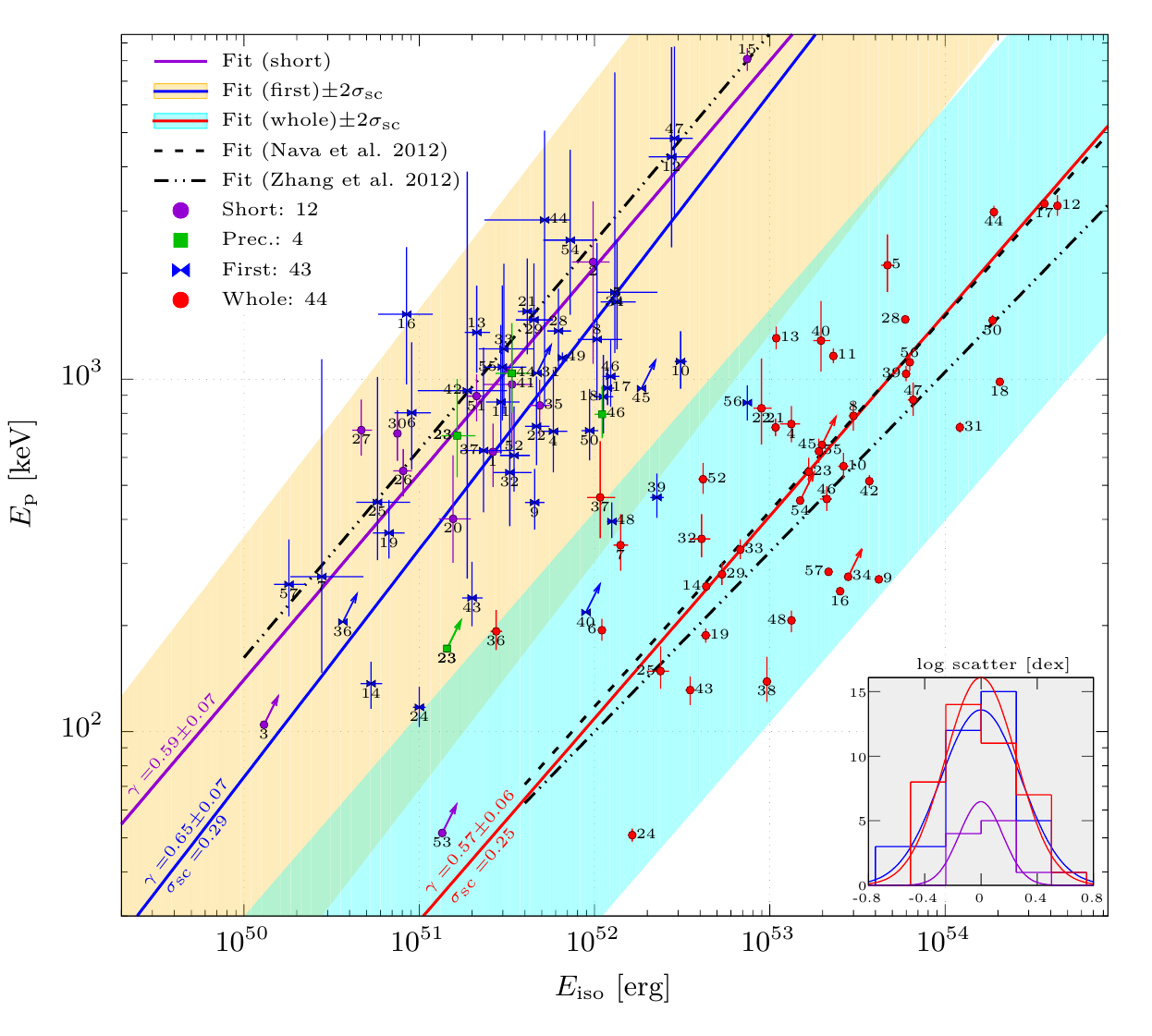}
  \caption{Rest frame $E_{\rm p}$--$E_{\rm iso}$ plane for all GRBs
    considered in this work.  The {\it short}, {\it first} and {\it
      whole} analysis results are shown with purple, blue and red
    symbols respectively, while the best fit correlations are shown
    with solid lines of the corresponding colors.  The numerical
    values of the slope and the scatter of the correlations are shown
    near the edges of the plots.  The inset plots show the histogram
    of the residuals from the best fit correlations.  The {\it first}
    analysis on precursors data are shown with green symbols.  The
    shaded areas are the 2$\sigma_{\rm sc}$ of the correlations of the
    {\it first} (orange) and {\it whole} (cyan) results.  Also shown
    are the $E_{\rm p}$--$E_{\rm iso}$ relations from the total sample
    of \citet{2012-Nava-Amati} (black dashed line) and from both the
    short and long GRB sample of
    \citet{2012-ZhangRevisitingLongShortClassif} (double dot--dashed
    lines) for comparison.}
  \label{fig-corr1}
\end{figure*}

\begin{figure*}
  \includegraphics[width=\textwidth]{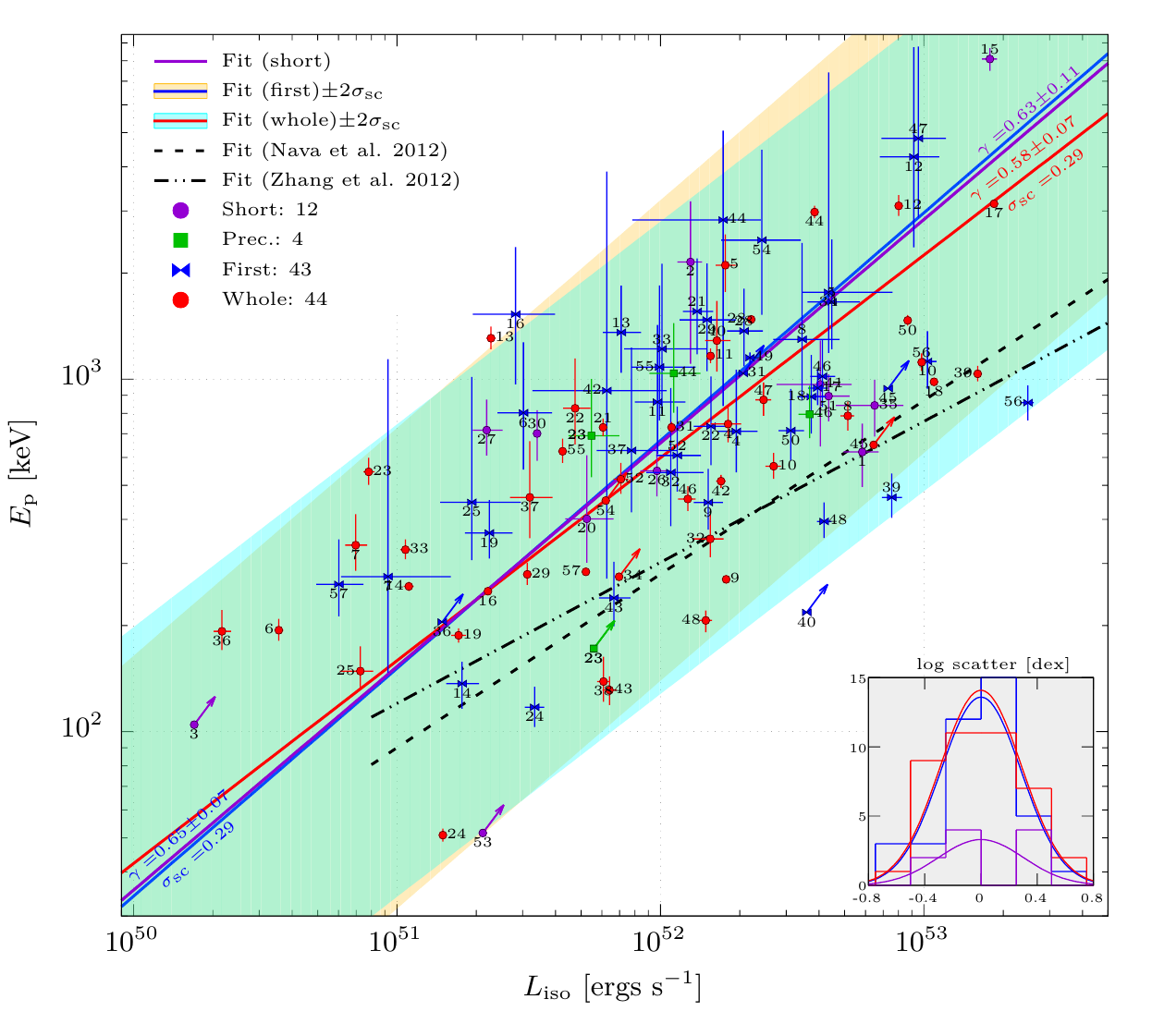}\\
  \caption{Rest frame $E_{\rm p}$--$L_{\rm iso}$ plane for all GRBs
    considered in this work with the same colors and symbols used in
    Fig. \ref{fig-corr1}.  Also shown are the $E_{\rm p}$--$L_{\rm
      p,iso}$ relations from the total sample of
    \citet{2012-Nava-Amati} (black dashed line) and from the combined
    short and long GRB sample of
    \citet{2012-ZhangRevisitingLongShortClassif} (double dot--dashed
    lines).}
  \label{fig-corr2}
\end{figure*}

In Fig. \ref{fig-corr1} we show the best fit correlations
(solid lines) on the $E_{\rm p}$--$E_{\rm iso}$ plane for the {\it
  short} (purple), {\it first} (blue) and {\it whole} (red) results,
as well as the histograms of residuals (inset plots).  For comparison
we also plot the corresponding relations from the total sample of
\citet{2012-Nava-Amati} (black dashed line) and from both the short
and long GRB sample of \citet{2012-ZhangRevisitingLongShortClassif}
(double dot--dashed lines).  In Fig. \ref{fig-corr2} we
show the corresponding results in the $E_{\rm p}$--$L_{\rm iso}$
plane.  For comparison we show the $E_{\rm p}$--$L_{\rm p,iso}$
relations from the total sample of \citet{2012-Nava-Amati} (black
dashed line) and from the combined short and long GRB sample of
\citet{2012-ZhangRevisitingLongShortClassif} (double dot--dashed
lines).

\begin{table*}
  \begin{center}
    \caption{Results of the statistical analysis of the $E_{\rm
        p}$--$E_{\rm iso}$ and $E_{\rm p}$--$L_{\rm iso}$ correlations
      for the {\em short}, {\em first} and {\em whole} results. $A$,
      $B$ and $\gamma$ are the correlation parameters, while
      $\rho_{\rm s} $ and P$_{\rm chance}$ are the Spearman's rank
      correlation coefficient and the associated chance probability.
      Results from precursors data are not considered.}
    \label{tab-corr}
    \begin{tabular}{ccccccccc}
    \hline
    \hline
    {\bf Correlation} &
    {\bf Results}     &
    {\bf No. GRBs}    &
    {\bf $A$}         &
    {\bf $B$}         &
    {\bf $\gamma$}    &
    {\bf $\sigma_{\rm sc}$} &
    {\bf $\rho_{\rm s}$}     &
    {\bf P$_{\rm chance}$}\\
    \hline
    \multirow{3}{*}{$\log \frac{E_{\rm p}}{\rm keV} = \gamma(\log \frac{E_{\rm iso}}{\rm erg} - A) + B$} 
    &   {\em short}   & 10 & 51.45 & 2.99 & 0.59 $\pm$ 0.07 & 0.15 & 0.71 & 0.02              \\
    &   {\em first}   & 39 & 51.62 & 2.92 & 0.65 $\pm$ 0.07 & 0.29 & 0.50 & $1\times 10^{-3}$ \\
    &   {\em whole}   & 41 & 53.21 & 2.73 & 0.57 $\pm$ 0.06 & 0.25 & 0.73 & $5\times 10^{-8}$ \\
    \hline
    \multirow{3}{*}{$\log \frac{E_{\rm p}}{\rm keV} = \gamma(\log \frac{L_{\rm iso}}{\rm erg\ s^{-1}} - A) + B$} 
    &   {\em short}   & 10 & 52.28 & 2.99 & 0.63 $\pm$ 0.11 & 0.30 & 0.50 & 0.14             \\
    &   {\em first}   & 39 & 52.15 & 2.92 & 0.65 $\pm$ 0.07 & 0.29 & 0.50 & $1\times 10^{-3}$ \\
    &   {\em whole}   & 41 & 51.92 & 2.73 & 0.58 $\pm$ 0.07 & 0.29 & 0.7  & $4\times 10^{-7}$ \\
    \hline
  \end{tabular}
  \end{center}
\end{table*}

\section{Discussion and conclusions}

We performed the spectral analysis of a sample long GRBs observed by
{\it Fermi}/GBM with a redshift estimate, using time integration equal
to 0.3 s rest frame ({\em first} analysis) and to the whole burst
duration ({\em whole} analysis).  Besides, we considered a sample of
short GRBs ({\em short} analysis), both by performing spectral
analysis of {\it Fermi}/GBM data and by reporting data from
\citet[][D14 sample]{2014-DAvanzo-SBAT8}. Our aim is to compare the
results of the {\em first} analysis to those of the {\em short} and
{\em whole} analysis.  The comparison of the relevant quantities are
shown in Fig. \ref{fig-histo}. Tab. \ref{tab-KS} shows the probability
that the distributions of the quantities shown in Fig. \ref{fig-histo}
are drawn from the same parent population.

The distributions of both $E_{\rm p}$ and $L_{\rm iso}$ are similar
for the the {\em short}, {\em first} and {\em whole} results.  The
distributions of low energy spectral index (\texttt{alpha}) for the
{\em short} and {\em first} results are very similar, but the
distribution for the {\it whole} results is significantly different
(K--S test probabilities $5.3 \times 10^{-5}$ and $\mathbf{8.9 \times
  10^{-4}}$, when compared to {\em first} and {\em short} results
respectively), with the latter showing lower values of \texttt{alpha}.
Also, the distribution of $E_{\rm iso}$ of the {\em whole} results is
significantly different from the {\em first} and {\em short} results
(K--S test probability: $4.1 \times 10^{-13}$ and $\mathbf{2.1 \times
  10^{-6}}$, when compared to {\em first} and {\em short} results
respectively), with the {\em whole} results lying a factor of a few
tens above the others.
\begin{table*}
  \begin{center}
    \caption{Results of Kolmogorov--Smirnov tests: column 2 and 3 show
      the probability that the distributions of the quantity shown in
      column 1 for the {\em short}, {\em first} and {\em whole}
      results are drawn from the same parent population.}
    \label{tab-KS}
    \begin{tabular}{cccc}
      \hline
      \hline
      {\bf Quantity}                          &
      {\bf {\em Short} vs. {\em First}}       &
      {\bf {\em Whole} vs. {\em First}}       &
      {\bf {\em Whole} vs. {\em Short}}       \\
      \hline
      alpha               & 0.83 & $5.3 \times 10^{-5}$   &  $8.9 \times 10^{-4}$\\
      log $E_{\rm p}$      & 0.71 & 0.11                  &   0.10\\
      log $E_{\rm iso}$    & 0.43 & $4.1 \times 10^{-13}$  &  $2.1 \times 10^{-6}$ \\
      log $L_{\rm iso}$    & 0.54 & 0.19                  &   0.29 \\
      \hline\hline
    \end{tabular}
  \end{center}
\end{table*}

The peak energy $E_{\rm p}$ of long GRBs, going from the first 0.3 s
(rest frame) to the whole burst duration, evolves either to lower or
higher energies, hence we do not find strong evidence for
hard--to--soft evolution of the peak energy.  The logarithmic median
value of the {\it whole} to {\it first} $E_{\rm p}$ ratio is
$\sim$~0.7 (Fig. \ref{fig-ratio}, left panel).  The total emitted
energy $E_{\rm iso}$ increases by a factor 5--10$^3$, with a
logarithmic median of $\sim$~35.  It is not clear what drives the
evolution of $E_{\rm p}$ towards either lower or higher energies,
since the $E_{\rm p, whole} / E_{\rm p, first}$ ratio does not show a
clear correlation with any other quantity.  The main driver for the
$E_{\rm iso}$ evolution is the total burst duration $\Delta T_{\rm
  whole,rest}$, i.e. longer burst likely evolve towards higher $E_{\rm
  iso}$ (Fig. \ref{fig-ratio}, right panel).

As discussed in \S\ref{sec-sample} the short GRB sample, although
relatively small when compared to the long sample, is a reasonably
good representation of the parent distribution of short GRBs
observable with currently available detectors.  Moreover, the short
GRB sample is large enough to provide evidence for significant
different distributions of \texttt{alpha} and $E_{\rm iso}$ when
compared to the {\it whole} sample.  Hence the K--S tests to compare
the spectral properties of the {\it short} and {\it first} samples are
reliable, and the corresponding distributions are actually
indistinguishable, i.e. we are unable to distinguish a short GRB from
the first 0.3 seconds of a long one with currently available
detectors.  Clearly, as new redshift estimates for SGRB become
available, our results may need to be reconsidered.

The plot of the $E_{\rm p}$--$E_{\rm iso}$ plane is shown in
Fig. \ref{fig-corr1}.  There is a clear correlation between $E_{\rm
  p}$ and $E_{\rm iso}$ for the {\em whole} results, with a chance
probability of obtaining a higher value of the Spearman's rank
correlation of $\sim 10^{-7}$ (Tab. \ref{tab-corr}).  In the $E_{\rm
  p}$--$E_{\rm iso}$ plane this is the well known Amati relation
\citep{2002-Amati-rel}.  The correlation slope and scatter
(0.57$\pm$0.06 and 0.25) are very similar to the ones found in
\citet{2012-Nava-Amati} for their total sample (0.55$\pm$0.02, 0.23),
and in \citet{2012-ZhangRevisitingLongShortClassif} for their long GRB
sample (0.51$\pm$0.03). For the {\em first} results we found a new
$E_{\rm p}$--$E_{\rm iso}$ relation with a probability $\sim 10^{-3}$
of being spurious.  The best fit {\em whole} relation lies at 3--4
$\sigma_{\rm sc}$ away from the {\em first} relation, hence the {\em
  first} and {\em whole} populations are well separated in the $E_{\rm
  p}$--$E_{\rm iso}$ plane.  The {\em short} GRBs alone do not provide
a strong statistical evidence for the existence of such a correlation
($P_{\rm chance} = 0.02$).  However, all {\em short} results lie
within 2$\sigma_{\rm sc}$ from the best fit relation for the {\em
  first} results.  Moreover, the best fit {\em short} correlation, if
it actually exists, lies very close to the {\em first} one, and
significantly away from the {\em whole} one.  Therefore, the {\em
  short} and {\em first} results are actually indistinguishable in the
$E_{\rm p}$--$E_{\rm iso}$ plane.  The lower limits for $E_{\rm p}$
and $E_{\rm iso}$ were not considered in the correlation analysis.
However, the true values of $E_{\rm p}$ and $E_{\rm iso}$ of the {\em
  short} and {\em first} population are not supposed to lie closer to
the {\em whole} correlation than their lower limits, as shown by the
arrows in Fig. \ref{fig-corr1}.  Hence our conclusions
can not be hampered by the presence of lower limits.  GRB precursors,
when present, also lie in the {\em short}--{\em first} region.

In the $E_{\rm p}$--$L_{\rm iso}$ (Fig. \ref{fig-corr2}) similar
considerations apply: there is a strong correlation for the {\em
  whole} results, a marginally significant correlation for the {\em
  first} results\footnote{The correlation analysis for the {\em first}
  results are the same in both the $E_{\rm p}$--$E_{\rm iso}$ and
  $E_{\rm p}$--$L_{\rm iso}$ planes, since the rest frame time
  interval is the same for all GRBs: 0.3 seconds.}, and a weak
correlation for the {\em short} results.  However, in the $E_{\rm
  p}$--$L_{\rm iso}$ plane all correlations overlap and are very
similar.  Note that these correlations are not equivalent to the
Yonetoku relation, since we used the $L_{\rm iso}$ values estimated on
the time averaged spectra, rather than the peak isotropic luminosity
$L_{\rm p,iso}$ \citep{2004-Yonetoku-EpLiso}.  Hence, we do not expect
to find the same results found in literature.  In particular we expect
our results to lie at lower $L_{\rm iso}$ since the peak luminosity is
by definition the highest luminosity for each burst.  Indeed the
Yonetoku relation found in \citet{2012-Nava-Amati} for their total
sample, and by \citet{2012-ZhangRevisitingLongShortClassif} for their
combined short and long sample, lie on the right of our best fit
correlation.  Nevertheless, our analysis shows that the $E_{\rm
  p}$--$L_{\rm iso}$ relation turns out to be very similar for the
{\em first} and {\em short} results (under the assumption that the
latter actually exists).  Hence, these correlations are possibly the
manifestation of the same physical process acting in all GRBs, and
even in small temporal intervals within a single GRB.

It has been debated if the Ep-Eiso and Ep-Liso correlation are
affected by selection effects
\citep{2005-Preece-TestingGRBRelationships,
  2005-Nakar-OutliersOf-EpEisoRelation,
  2007-Butler-CompleteCatalogSwift-SpectraDuration,
  2009-Butler-GeneralizedTestSelectionEffects-Correlations,
  2011-Shahmoradi-ImpactOfDetectorThreshold,
  2012-Kocevski-OriginOfHighEnergyCorrelations}. The possible
existence of similar correlation within individual GRBs, i.e. between
the peak energy and the luminosity as a function of time in a single
GRB \citep{2009-Firmani-TimeResolvedCorrel,
  2010-Ghirlanda-TimeResolvedFermi, 2011-Ghirlanda-ShortVsLong,
  2011-Ghirlanda-SpectralEvolutionFermiSGRB,
  2012-Frontera-TimeResolvCorr} seems to point to a physical origin of
these correlations. Similarly the use of a flux limited complete
sample of long GRBs seems to support the idea that instrumental
selection effects are not shaping these correlations
\citep{2012-Nava-Amati, 2012-Ghirlanda-EpLisoSelectionEffects}.  Hence
these correlations are likely the manifestation of fundamental GRB
properties.  Since our correlations for the {\it whole} analysis
(Tab. \ref{tab-corr}) are very similar to those found in literature we
do not expect our long sample to be strongly biased by selection
effects.  As a consequence, also the {\it first} correlation is not
biased, since the sample is the same.  The {\it short} analysis, on
the other hand, has been performed on a flux limited short GRB sample
with a redshift completeness of $\sim$~60\%, hence we do not expect
the selection effects (beyond the limiting flux threshold) to play a
dominant role.

The 0.3 s (rest frame) time scale chosen for the {\it first} analysis
has a clear interpretation: it is the representative duration of the
short GRBs in the D14 catalog.  Since the main driver for the $E_{\rm
  iso}$ evolution is the integration time (Fig. \ref{fig-ratio}, right
panel), a longer timescale would result in higher values of $E_{\rm
  iso,first}$.  Hence, in order to obtain significantly higher (or
lower) values of $E_{\rm iso}$, we should overcome the intrinsic
scatter of the correlations, namely 0.25--0.3 dex (a factor $\sim$~2,
i.e. $\Delta T_{\rm first} \lesssim$~0.15 s or $\Delta T_{\rm first}
\gtrsim$~0.6 s).

In summary, we found that the intrinsic spectral properties (peak
energy and luminosity) of both the short GRBs and the first 0.3 s
(rest frame) of long ones are actually indistinguishable.  Hence if
the central engine of a long GRB would stop working after $\sim$~0.3
s, we would have no means to distinguish it from a genuine short GRB.
Clearly, short and long GRBs remains two distinct phenomena, each one
with its own duration.  In particular short GRBs lasts longer or
shorter than 0.3 s.  Likewise, we do not expect the short--like phase
at the beginning of long GRBs to always lasts 0.3 s.  Our findings are
in agreement with those in \citet{2009-Ghirlanda-ShortVsLong}, which
found no differences in the (observed) spectral properties of short
GRBs and the first 1--2 s of long GRBs.  We extended this work by
comparing the intrinsic (rest frame) properties rather than the
observed ones.

Moreover, we found that the spectral quantities in the first 0.3 s of
long GRBs define new $E_{\rm p}$--$E_{\rm iso}$ and $E_{\rm
  p}$--$L_{\rm iso}$ correlations.  These correlations are possibly
the manifestation of an underlying physical process common to all
GRBs, despite the possibly different progenitors of short and long
GRBs, and the great variety of energetics and spectral properties
involved.

\bigskip

\noindent 
{\bf ACKNOWLEDGMENTS}

\noindent

We gratefully acknowledge the Referee for the useful comments and
suggestions which greatly improved the paper.  The research activity
of A. Melandri, M.G. Bernardini and P. D'Avanzo is supported by ASI
grant INAF I/004/11/1.  G. Calderone was supported by PRIN INAF 2011
(1.05.01.09.15).

%
\bibliographystyle{mn2e}

\appendix

\section{Results of spectral analysis}
%
\begin{table*}
  \begin{center}
    \caption{List of GRBs considered in this work.  Columns are: [1]
      GRB identifier; [2] GRB name [3] fraction of day of trigger
      time; [4] redshift; [5] $T_{90}$ duration in the {\it Fermi}/GBM
      50--300 keV energy range (observer frame) {\it Fermi}/GBM
      \citep{2014-vonKienlin-GBMCat-4yr}; [6] list of detectors used
      for spectral analysis; [7] background time selection; [8]
      redshift reference.  Rows with missing data refer to short GRBs
      in the D14 sample \citep{2014-DAvanzo-SBAT8}.}
    \label{tab1}
    \renewcommand{\arraystretch}{1.5}
    \tiny
    \begin{tabular}{clclcccrr}
      \hline
      \hline
      {\bf ID}                          &
      {\bf GRB }                        &
      {\bf (day frac.)}                 &
      {\bf Redshift}                    &
      {\bf T$_{90}$ [s]}                 &
      {\bf Detectors}                   &
      {\bf Background time sel. [s]}    &
      {\bf Redshift ref.}               &\\
 \hline
  1  &   051221A  &        &   0.5465  &           &                   &                              &                     &   \\
  2  &   070714B  &        &     0.92  &           &                   &                              &                     &   \\
  3  &    080123  &        &    0.495  &           &                   &                              &                     &   \\
  4  &    080804  &   972  &   2.2045  &   24.704  &       n6, n7, b1  &            -100:-10, 50:200  &    \citet{z080804}  &   \\
  5  &    080810  &   549  &     3.35  &   107.46  &   n7, n8, nb, b1  &           -100:-20, 110:200  &    \citet{z080810}  &   \\
  6  &   080916A  &   406  &    0.689  &   46.337  &       n7, n8, b1  &           -100:-20, 100:200  &   \citet{z080916A}  &   \\
  7  &    081109  &   293  &   0.9787  &   58.369  &       n9, na, b1  &            -100:-20, 40:200  &    \citet{z081109}  &   \\
  8  &    081121  &   858  &    2.512  &   41.985  &       na, nb, b1  &            -100:-10, 50:200  &    \citet{z081121}  &   \\
  9  &    081221  &   681  &     2.26  &   29.697  &       n1, n2, b0  &           -100:-10, 100:200  &    \citet{z081221}  &   \\
 10  &    081222  &   204  &     2.77  &    18.88  &       n0, n1, b0  &            -200:-10, 50:200  &    \citet{z081222}  &   \\
 11  &    090102  &   122  &    1.547  &   26.624  &       na, nb, b1  &            -100:-20, 50:200  &    \citet{z090102}  &   \\
 12  &    090323  &   002  &     3.57  &   135.17  &       n9, nb, b1  &           -100:-20, 200:300  &    \citet{z090323}  &   \\
 13  &    090328  &   401  &    0.736  &   61.697  &       n6, n7, b1  &           -100:-10, 100:200  &    \citet{z090328}  &   \\
 14  &    090424  &   592  &    0.544  &   14.144  &   n7, n8, nb, b1  &            -100:-5, 100:200  &    \citet{z090424}  &   \\
 15  &    090510  &        &    0.903  &           &                   &                              &                     &   \\
 16  &    090618  &   353  &     0.54  &   112.39  &           n4, b0  &           -100:-20, 200:400  &    \citet{z090618}  &   \\
 17  &    090902  &   462  &    1.822  &   19.328  &       n0, n1, b0  &            -100:-10, 60:150  &    \citet{z090902}  &   \\
 18  &   090926A  &   181  &   2.1062  &    13.76  &       n6, n7, b1  &             -55:-10, 60:200  &   \citet{z090926A}  &   \\
 19  &   090926B  &   914  &     1.24  &   55.553  &   n7, n8, nb, b1  &             -100:-5, 60:200  &   \citet{z090926B}  &   \\
 20  &    090927  &   422  &     1.37  &    0.512  &   n2, n9, na, b1  &            -100:-20, 10:200  &    \citet{z090927}  &   \\
 21  &    091003  &   191  &   0.8969  &   20.224  &       n3, n6, b1  &            -100:-10, 50:200  &    \citet{z091003}  &   \\
 22  &   091020A  &   900  &     1.71  &   24.256  &       n2, n5, b0  &            -100:-15, 50:200  &   \citet{z091020A}  &   \\
 23  &    091024  &   372  &    1.092  &   93.954  &       n7, n8, b1  &            -100:-10, 60:200  &    \citet{z091024}  &   \\
 23  &    091024  &   380  &    1.092  &   93.954  &       n6, n9, b1  &   -200:-40, 60:200, 500:700  &    \citet{z091024}  &   \\
 24  &    091127  &   976  &     0.49  &    8.701  &   n6, n7, n9, b1  &             -100:-5, 30:200  &    \citet{z091127}  &   \\
 25  &   091208B  &   410  &    1.063  &    12.48  &       n9, na, b1  &           -100:-20, 130:250  &   \citet{z091208B}  &   \\
 26  &   100117A  &        &     0.92  &           &                   &                              &                     &   \\
 27  &    100206  &   563  &   0.4068  &    0.128  &       n0, n3, b0  &              -100:-5, 5:200  &    \citet{z100206}  &   \\
 28  &    100414  &   097  &    1.368  &   26.497  &   n7, n9, nb, b1  &           -100:-20, 110:200  &    \citet{z100414}  &   \\
 29  &   100615A  &   083  &    1.398  &   37.377  &   n6, n7, n8, b1  &            -100:-5, 100:200  &   \citet{z100615A}  &   \\
 30  &   100625A  &        &    0.452  &           &                   &                              &                     &   \\
 31  &   100728A  &   095  &    1.567  &   165.38  &       n0, n1, b0  &           -100:-20, 300:500  &   \citet{z100728A}  &   \\
 32  &   100728B  &   439  &    2.106  &    10.24  &   n6, n7, n8, b1  &            -100:-10, 10:200  &   \citet{z100728B}  &   \\
 33  &   100814A  &   160  &     1.44  &   150.53  &       n7, n8, b1  &   -100:-20, 90:120, 200:300  &   \citet{z100814A}  &   \\
 34  &   100906A  &   576  &    1.727  &   110.59  &           nb, b1  &           -100:-10, 150:250  &   \citet{z100906A}  &   \\
 35  &   101219A  &        &    0.718  &           &                   &                              &                     &   \\
 36  &   110106B  &   893  &    0.618  &   35.521  &   n9, na, nb, b1  &            -100:-20, 40:200  &   \citet{z110106B}  &   \\
 37  &   110128A  &   073  &    2.339  &    12.16  &   n6, n7, n9, b1  &            -100:-10, 10:200  &   \citet{z110128A}  &   \\
 38  &   110213A  &   220  &     1.46  &   34.305  &       n3, n4, b0  &            -100:-10, 50:200  &   \citet{z110213A}  &   \\
 39  &   110731A  &   465  &     2.83  &    7.485  &       n0, n3, b0  &             -100:-5, 20:200  &   \citet{z110731A}  &   \\
 40  &   110818A  &   860  &     3.36  &   67.073  &   n7, n8, nb, b1  &            -100:-20, 50:200  &   \citet{z110818A}  &   \\
 41  &   111117A  &        &      1.3  &           &                   &                              &                     &   \\
 42  &   120119A  &   170  &    1.728  &   55.297  &       n9, nb, b1  &           -100:-20, 100:200  &   \citet{z120119A}  &   \\
 43  &   120326A  &   056  &    1.798  &   11.776  &   n0, n1, n2, b0  &            -100:-10, 20:200  &   \citet{z120326A}  &   \\
 44  &   120711A  &   115  &    1.405  &   44.033  &       n2, na, b0  &    -100:-20, 10:50, 150:250  &   \citet{z120711A}  &   \\
 45  &   120712A  &   571  &   4.1745  &   22.528  &       n6, n7, b1  &            -100:-10, 30:200  &   \citet{z120712A}  &   \\
 46  &   120716A  &   712  &    2.486  &    234.5  &       n9, na, b1  &   -100:-10, 20:160, 250:400  &   \citet{z120716A}  &   \\
 47  &   120909A  &   070  &     3.93  &   112.07  &       n7, n8, b1  &           -100:-10, 150:300  &   \citet{z120909A}  &   \\
 48  &   121128A  &   212  &      2.2  &   17.344  &       n3, n4, b0  &             -100:-5, 50:200  &   \citet{z121128A}  &   \\
 49  &   130427A  &   324  &   0.3399  &   138.24  &       n9, na, b1  &                    -200:-10  &   \citet{z130427A}  &   \\
 50  &   130518A  &   580  &    2.488  &   48.577  &   n3, n7, b0, b1  &           -100:-20, 100:200  &   \citet{z130518A}  &   \\
 51  &   130603B  &        &    0.356  &           &                   &                              &                     &   \\
 52  &   130610A  &   133  &    2.092  &    21.76  &       n7, n8, b1  &            -100:-40, 30:200  &   \citet{z130610A}  &   \\
 53  &   131004A  &   904  &    0.717  &    1.152  &       n9, na, b1  &             -100:-5, 10:200  &   \citet{z131004A}  &   \\
 54  &   131011A  &   741  &    1.874  &   77.057  &   n7, n9, nb, b1  &           -100:-20, 100:200  &   \citet{z131011A}  &   \\
 55  &   131105A  &   087  &    1.686  &   112.64  &       n6, n7, b1  &           -200:-20, 150:300  &   \citet{z131105A}  &   \\
 56  &   131108A  &   862  &      2.4  &   18.496  &       n0, n3, b0  &            -200:-10, 50:200  &   \citet{z131108A}  &   \\
 57  &   131231A  &   198  &    0.642  &   31.232  &       n0, n3, b1  &           -100:-20, 100:200  &   \citet{z131231A}  &   \\
      \hline\hline
    \end{tabular}
  \end{center}
\end{table*}

\begin{table*}
  \begin{center}
    \caption{List of intrinsic (rest frame) spectral quantities for
      our GRB sample.  Columns are: [1] GRB identifier; [2] GRB name;
      [3] Time selection (observer frame) [4] analysis specification;
      [5] spectral model; [6] \texttt{alpha} spectral index; [7]
      \texttt{beta} spectral index spectral; [8] log $E_{\rm p}$; [9]
      log $E_{\rm iso}$ in the 1 kev--10 MeV energy range; [10] value
      of the Cash fit statistic and [11] degrees of freedom.}
    \label{tab2}
    \renewcommand{\arraystretch}{1.7}
    \tiny
    \begin{tabular}{rlccccccccccc}
      \hline
      \hline
      {\bf ID}                        &
      {\bf GRB}                       &
      \multicolumn{2}{c}{{\bf Time sel. [s]}}  &
      {\bf Analysis}                  &
      {\bf Model}                     &
      {\bf Alpha}                     &
      {\bf Beta}                      &
      {\bf $\log E_{\rm p}$ [keV]}     &
      {\bf $\log E_{\rm iso}$ [erg]}   &
      {\bf C--STAT}                   &
      {\bf DOF}                       &\\
      \hline
  \multirow{1}{*}{1}  &   \multirow{1}{*}{051221A}  &            &           &   (short D14)  &         &         $ -1.08\,^{+0.13}_{-0.13} $  &                                &      $ 2.793\,^{+0.081}_{-0.099} $  &      $ 51.420\,^{+0.051}_{-0.058} $  &           &     0  &   \\ \hline
  \multirow{1}{*}{2}  &   \multirow{1}{*}{070714B}  &            &           &   (short D14)  &         &         $ -0.86\,^{+0.10}_{-0.10} $  &                                &         $ 3.33\,^{+0.17}_{-0.29} $  &      $ 51.991\,^{+0.094}_{-0.121} $  &           &     0  &   \\ \hline
  \multirow{1}{*}{3}  &    \multirow{1}{*}{080123}  &            &           &   (short D14)  &         &         $ -1.20\,^{+0.38}_{-0.38} $  &                                &                          $> 2.02 $  &                          $> 50.11 $  &           &     0  &   \\ \hline
  \multirow{2}{*}{4}  &    \multirow{2}{*}{080804}  &      -0.6  &    0.361  &       (first)  &   Band  &         $ -0.25\,^{+0.64}_{-0.44} $  &                      -2.3$^a$  &         $ 2.85\,^{+0.18}_{-0.12} $  &      $ 51.765\,^{+0.080}_{-0.069} $  &   408.58  &   362  &          \\
                      &                             &    -1.024  &   22.528  &       (whole)  &   Band  &      $ -0.669\,^{+0.089}_{-0.078} $  &      $ -2.5\,^{+0.3}_{-1.1} $  &      $ 2.873\,^{+0.051}_{-0.052} $  &      $ 53.123\,^{+0.049}_{-0.065} $  &   510.38  &   361  &   \\ \hline
  \multirow{2}{*}{5}  &    \multirow{2}{*}{080810}  &        -1  &    0.305  &       (first)  &    CPL  &         $ -0.71\,^{+0.29}_{-0.23} $  &                                &         $ 3.22\,^{+0.18}_{-0.13} $  &      $ 52.128\,^{+0.107}_{-0.093} $  &   501.83  &   482  &          \\
                      &                             &    -10.24  &   105.47  &       (whole)  &   Band  &      $ -1.090\,^{+0.063}_{-0.059} $  &                      -2.3$^a$  &      $ 3.323\,^{+0.087}_{-0.076} $  &      $ 53.671\,^{+0.039}_{-0.037} $  &     1221  &   482  &   \\ \hline
  \multirow{2}{*}{6}  &   \multirow{2}{*}{080916A}  &      -0.2  &    0.307  &       (first)  &   Band  &         $ -0.29\,^{+0.45}_{-0.32} $  &                      -2.3$^a$  &         $ 2.90\,^{+0.20}_{-0.16} $  &      $ 50.957\,^{+0.109}_{-0.099} $  &   377.97  &   361  &          \\
                      &                             &    -1.024  &     51.2  &       (whole)  &   Band  &      $ -1.047\,^{+0.067}_{-0.065} $  &                      -2.3$^a$  &      $ 2.288\,^{+0.033}_{-0.030} $  &      $ 52.042\,^{+0.017}_{-0.017} $  &   683.35  &   361  &   \\ \hline
  \multirow{2}{*}{7}  &    \multirow{2}{*}{081109}  &      -1.2  &   -0.606  &       (first)  &    CPL  &         $ -1.20\,^{+0.75}_{-0.49} $  &                                &         $ 2.44\,^{+0.62}_{-0.27} $  &         $ 50.44\,^{+0.24}_{-0.18} $  &   420.97  &   360  &          \\
                      &                             &     -5.12  &   34.816  &       (whole)  &    CPL  &         $ -1.22\,^{+0.12}_{-0.11} $  &                                &      $ 2.529\,^{+0.088}_{-0.073} $  &      $ 52.148\,^{+0.043}_{-0.039} $  &   573.17  &   360  &   \\ \hline
  \multirow{2}{*}{8}  &    \multirow{2}{*}{081121}  &       0.8  &    1.854  &       (first)  &    CPL  &         $ -0.88\,^{+0.37}_{-0.28} $  &                                &         $ 3.11\,^{+0.27}_{-0.17} $  &         $ 52.01\,^{+0.14}_{-0.11} $  &   383.75  &   359  &          \\
                      &                             &         0  &    20.48  &       (whole)  &   Band  &      $ -0.682\,^{+0.087}_{-0.078} $  &      $ -2.2\,^{+0.1}_{-0.2} $  &      $ 2.896\,^{+0.040}_{-0.042} $  &      $ 53.477\,^{+0.025}_{-0.027} $  &   439.43  &   358  &   \\ \hline
  \multirow{2}{*}{9}  &    \multirow{2}{*}{081221}  &      -0.8  &    0.178  &       (first)  &   Band  &          $ 0.92\,^{+1.17}_{-0.75} $  &                      -2.3$^a$  &      $ 2.650\,^{+0.095}_{-0.077} $  &      $ 51.659\,^{+0.056}_{-0.055} $  &   368.91  &   362  &          \\
                      &                             &    -1.024  &   75.776  &       (whole)  &   Band  &      $ -0.858\,^{+0.029}_{-0.027} $  &      $ -3.3\,^{+0.2}_{-0.2} $  &   $ 2.4321\,^{+0.0083}_{-0.0089} $  &   $ 53.6217\,^{+0.0101}_{-0.0093} $  &   894.98  &   361  &   \\ \hline
 \multirow{2}{*}{10}  &    \multirow{2}{*}{081222}  &         0  &    1.131  &       (first)  &   Band  &         $ -0.79\,^{+0.12}_{-0.11} $  &                      -2.3$^a$  &      $ 3.050\,^{+0.086}_{-0.077} $  &      $ 52.490\,^{+0.034}_{-0.033} $  &   421.53  &   361  &          \\
                      &                             &    -1.024  &    35.84  &       (whole)  &   Band  &      $ -0.886\,^{+0.063}_{-0.059} $  &      $ -2.4\,^{+0.2}_{-0.2} $  &      $ 2.753\,^{+0.038}_{-0.036} $  &      $ 53.420\,^{+0.028}_{-0.031} $  &   429.81  &   360  &   \\ \hline
 \multirow{2}{*}{11}  &    \multirow{2}{*}{090102}  &         0  &    0.764  &       (first)  &   Band  &         $ -0.69\,^{+0.36}_{-0.28} $  &                      -2.3$^a$  &         $ 2.94\,^{+0.22}_{-0.15} $  &      $ 51.466\,^{+0.106}_{-0.086} $  &   372.54  &   359  &          \\
                      &                             &    -1.792  &   36.096  &       (whole)  &    CPL  &      $ -0.966\,^{+0.022}_{-0.021} $  &                                &      $ 3.066\,^{+0.021}_{-0.020} $  &      $ 53.363\,^{+0.012}_{-0.012} $  &   553.16  &   359  &   \\ \hline
 \multirow{2}{*}{12}  &    \multirow{2}{*}{090323}  &       1.4  &    2.771  &       (first)  &    CPL  &         $ -1.04\,^{+0.19}_{-0.14} $  &                                &         $ 3.63\,^{+0.31}_{-0.26} $  &      $ 52.439\,^{+0.096}_{-0.129} $  &   372.58  &   360  &          \\
                      &                             &    -3.072  &   245.76  &       (whole)  &   Band  &      $ -1.255\,^{+0.012}_{-0.012} $  &                      -2.3$^a$  &      $ 3.492\,^{+0.030}_{-0.029} $  &      $ 54.641\,^{+0.010}_{-0.010} $  &   1596.9  &   360  &   \\ \hline
 \multirow{2}{*}{13}  &    \multirow{2}{*}{090328}  &       3.8  &    4.321  &       (first)  &    CPL  &      $ -1.009\,^{+0.102}_{-0.091} $  &                                &         $ 3.13\,^{+0.13}_{-0.11} $  &      $ 51.328\,^{+0.076}_{-0.069} $  &   433.47  &   361  &          \\
                      &                             &    -4.096  &   78.848  &       (whole)  &    CPL  &      $ -1.140\,^{+0.019}_{-0.019} $  &                                &      $ 3.116\,^{+0.033}_{-0.031} $  &      $ 53.036\,^{+0.018}_{-0.017} $  &   754.48  &   361  &   \\ \hline
 \multirow{2}{*}{14}  &    \multirow{2}{*}{090424}  &      -0.2  &    0.263  &       (first)  &   Band  &         $ -0.90\,^{+0.20}_{-0.16} $  &      $ -2.6\,^{+0.3}_{-0.6} $  &      $ 2.136\,^{+0.062}_{-0.071} $  &      $ 50.724\,^{+0.064}_{-0.060} $  &   574.44  &   480  &          \\
                      &                             &    -1.024  &   59.392  &       (whole)  &   Band  &      $ -1.060\,^{+0.017}_{-0.017} $  &      $ -2.8\,^{+0.2}_{-0.3} $  &      $ 2.412\,^{+0.013}_{-0.013} $  &      $ 52.637\,^{+0.018}_{-0.020} $  &   951.52  &   480  &   \\ \hline
 \multirow{1}{*}{15}  &    \multirow{1}{*}{090510}  &            &           &   (short D14)  &         &      $ -0.820\,^{+0.020}_{-0.020} $  &      $ -2.8\,^{+0.3}_{-0.3} $  &      $ 3.908\,^{+0.031}_{-0.033} $  &      $ 52.871\,^{+0.018}_{-0.019} $  &           &     0  &   \\ \hline
 \multirow{2}{*}{16}  &    \multirow{2}{*}{090618}  &      -0.8  &   -0.338  &       (first)  &    CPL  &         $ -0.29\,^{+0.66}_{-0.35} $  &                                &         $ 3.18\,^{+0.19}_{-0.20} $  &         $ 50.93\,^{+0.15}_{-0.16} $  &   245.72  &   240  &          \\
                      &                             &    -1.024  &   174.08  &       (whole)  &   Band  &      $ -1.166\,^{+0.013}_{-0.012} $  &   $ -2.51\,^{+0.05}_{-0.05} $  &   $ 2.3981\,^{+0.0097}_{-0.0100} $  &   $ 53.4013\,^{+0.0074}_{-0.0075} $  &   1493.4  &   239  &   \\ \hline
 \multirow{2}{*}{17}  &    \multirow{2}{*}{090902}  &      -0.4  &    0.447  &       (first)  &   Band  &         $ -0.29\,^{+0.17}_{-0.15} $  &                      -2.3$^a$  &      $ 2.974\,^{+0.055}_{-0.048} $  &      $ 52.074\,^{+0.039}_{-0.036} $  &    376.4  &   362  &          \\
                      &                             &    -1.024  &   55.296  &       (whole)  &   Band  &   $ -1.0214\,^{+0.0050}_{-0.0047} $  &                      -2.3$^a$  &   $ 3.4976\,^{+0.0079}_{-0.0082} $  &   $ 54.5667\,^{+0.0038}_{-0.0040} $  &   1879.8  &   362  &   \\ \hline
 \multirow{2}{*}{18}  &   \multirow{2}{*}{090926A}  &         0  &    0.932  &       (first)  &   Band  &         $ -0.69\,^{+0.21}_{-0.17} $  &                      -2.3$^a$  &         $ 2.95\,^{+0.12}_{-0.10} $  &      $ 52.050\,^{+0.057}_{-0.053} $  &   391.41  &   362  &          \\
                      &                             &    -7.168  &     51.2  &       (whole)  &   Band  &      $ -0.821\,^{+0.010}_{-0.010} $  &   $ -2.49\,^{+0.06}_{-0.07} $  &   $ 2.9924\,^{+0.0082}_{-0.0082} $  &   $ 54.3125\,^{+0.0061}_{-0.0064} $  &   973.28  &   361  &   \\ \hline
 \multirow{2}{*}{19}  &   \multirow{2}{*}{090926B}  &       0.6  &    1.272  &       (first)  &    CPL  &             $ 1.2\,^{+2.0}_{-1.1} $  &                                &      $ 2.564\,^{+0.093}_{-0.073} $  &      $ 50.827\,^{+0.089}_{-0.092} $  &   527.49  &   482  &          \\
                      &                             &    -1.024  &   55.296  &       (whole)  &   Band  &          $ 0.16\,^{+0.15}_{-0.13} $  &      $ -2.9\,^{+0.2}_{-0.3} $  &      $ 2.273\,^{+0.020}_{-0.020} $  &      $ 52.634\,^{+0.028}_{-0.028} $  &   684.57  &   481  &   \\ \hline
 \multirow{1}{*}{20}  &    \multirow{1}{*}{090927}  &    -0.256  &    0.448  &       (short)  &    CPL  &         $ -0.80\,^{+0.37}_{-0.30} $  &                                &         $ 2.60\,^{+0.18}_{-0.12} $  &      $ 51.193\,^{+0.101}_{-0.082} $  &   523.39  &   482  &   \\ \hline
 \multirow{2}{*}{21}  &    \multirow{2}{*}{091003}  &         0  &    0.569  &       (first)  &   Band  &      $ -1.112\,^{+0.082}_{-0.075} $  &                      -2.3$^a$  &         $ 3.19\,^{+0.15}_{-0.12} $  &      $ 51.616\,^{+0.061}_{-0.054} $  &   382.73  &   362  &          \\
                      &                             &    -1.024  &   32.768  &       (whole)  &   Band  &      $ -1.068\,^{+0.022}_{-0.021} $  &                      -2.3$^a$  &      $ 2.863\,^{+0.025}_{-0.024} $  &   $ 53.0331\,^{+0.0092}_{-0.0090} $  &   501.81  &   362  &   \\ \hline
 \multirow{2}{*}{22}  &   \multirow{2}{*}{091020A}  &      -0.2  &    0.613  &       (first)  &    CPL  &         $ -1.08\,^{+0.17}_{-0.15} $  &                                &         $ 2.87\,^{+0.14}_{-0.11} $  &      $ 51.668\,^{+0.074}_{-0.065} $  &    381.4  &   362  &          \\
                      &                             &    -6.144  &   45.056  &       (whole)  &    CPL  &      $ -1.386\,^{+0.072}_{-0.069} $  &                                &         $ 2.92\,^{+0.14}_{-0.10} $  &      $ 52.952\,^{+0.058}_{-0.045} $  &   473.64  &   362  &   \\ \hline
 \multirow{3}{*}{23}  &    \multirow{3}{*}{091024}  &       2.4  &    3.028  &       (prec.)  &    CPL  &         $ -0.62\,^{+0.30}_{-0.25} $  &                                &         $ 2.84\,^{+0.16}_{-0.12} $  &      $ 51.216\,^{+0.106}_{-0.088} $  &   390.51  &   360  &          \\
                      &                             &         5  &    5.628  &       (prec.)  &   Band  &         $ -0.80\,^{+0.86}_{-0.44} $  &                      -2.3$^a$  &                          $> 2.24 $  &                          $> 51.16 $  &   381.46  &   363  &          \\
                      &                             &   -16.384  &   430.09  &       (whole)  &    CPL  &      $ -1.045\,^{+0.043}_{-0.042} $  &                                &      $ 2.737\,^{+0.040}_{-0.037} $  &      $ 53.222\,^{+0.022}_{-0.021} $  &   1134.7  &   363  &   \\ \hline
 \multirow{2}{*}{24}  &    \multirow{2}{*}{091127}  &      -0.2  &    0.247  &       (first)  &   Band  &         $ -0.56\,^{+0.20}_{-0.17} $  &   $ -2.11\,^{+0.07}_{-0.09} $  &      $ 2.069\,^{+0.058}_{-0.056} $  &      $ 51.000\,^{+0.036}_{-0.037} $  &   492.47  &   481  &          \\
                      &                             &    -1.024  &    15.36  &       (whole)  &   Band  &      $ -1.252\,^{+0.067}_{-0.063} $  &   $ -2.21\,^{+0.02}_{-0.02} $  &      $ 1.706\,^{+0.018}_{-0.018} $  &   $ 52.2153\,^{+0.0076}_{-0.0076} $  &   713.48  &   481  &   \\ \hline
 \multirow{2}{*}{25}  &   \multirow{2}{*}{091208B}  &      -0.6  &    0.019  &       (first)  &    CPL  &         $ -0.83\,^{+0.59}_{-0.43} $  &                                &         $ 2.65\,^{+0.36}_{-0.16} $  &         $ 50.76\,^{+0.19}_{-0.12} $  &   415.97  &   359  &          \\
                      &                             &    -1.024  &    66.56  &       (whole)  &   Band  &         $ -0.92\,^{+0.17}_{-0.17} $  &      $ -2.6\,^{+0.2}_{-2.5} $  &      $ 2.172\,^{+0.071}_{-0.050} $  &      $ 52.376\,^{+0.049}_{-0.077} $  &   837.99  &   358  &   \\ \hline
 \multirow{1}{*}{26}  &   \multirow{1}{*}{100117A}  &            &           &   (short D14)  &         &         $ -0.15\,^{+0.21}_{-0.21} $  &                                &      $ 2.740\,^{+0.062}_{-0.073} $  &      $ 50.908\,^{+0.051}_{-0.057} $  &           &     0  &   \\ \hline
 \multirow{1}{*}{27}  &    \multirow{1}{*}{100206}  &      -0.1  &      0.2  &       (short)  &    CPL  &         $ -0.42\,^{+0.17}_{-0.15} $  &                                &      $ 2.855\,^{+0.087}_{-0.072} $  &      $ 50.669\,^{+0.060}_{-0.053} $  &   399.21  &   362  &   \\ \hline
 \multirow{2}{*}{28}  &    \multirow{2}{*}{100414}  &       0.4  &     1.11  &       (first)  &   Band  &         $ -0.42\,^{+0.23}_{-0.19} $  &                      -2.3$^a$  &         $ 3.14\,^{+0.12}_{-0.10} $  &      $ 51.794\,^{+0.072}_{-0.064} $  &   461.74  &   479  &          \\
                      &                             &    -1.024  &   62.464  &       (whole)  &   Band  &      $ -0.594\,^{+0.017}_{-0.017} $  &      $ -3.5\,^{+0.4}_{-0.7} $  &      $ 3.169\,^{+0.012}_{-0.012} $  &      $ 53.773\,^{+0.012}_{-0.013} $  &   1109.4  &   478  &   \\ \hline
 \multirow{2}{*}{29}  &   \multirow{2}{*}{100615A}  &      -0.2  &    0.519  &       (first)  &    CPL  &         $ -0.82\,^{+0.15}_{-0.13} $  &                                &         $ 3.17\,^{+0.16}_{-0.15} $  &         $ 51.65\,^{+0.11}_{-0.10} $  &   498.68  &   485  &          \\
                      &                             &    -1.024  &   39.936  &       (whole)  &    CPL  &      $ -1.343\,^{+0.044}_{-0.043} $  &                                &      $ 2.446\,^{+0.033}_{-0.030} $  &      $ 52.727\,^{+0.016}_{-0.015} $  &   783.08  &   485  &   \\ \hline
 \multirow{1}{*}{30}  &   \multirow{1}{*}{100625A}  &            &           &   (short D14)  &         &         $ -0.60\,^{+0.11}_{-0.11} $  &                                &      $ 2.846\,^{+0.066}_{-0.078} $  &      $ 50.875\,^{+0.017}_{-0.018} $  &           &     0  &   \\ \hline
 \multirow{2}{*}{31}  &   \multirow{2}{*}{100728A}  &       3.6  &     4.37  &       (first)  &   Band  &         $ -0.54\,^{+0.57}_{-0.35} $  &                      -2.3$^a$  &                          $> 3.02 $  &                          $> 51.67 $  &   381.15  &   362  &          \\
                      &                             &    -4.096  &   278.53  &       (whole)  &   Band  &      $ -0.730\,^{+0.022}_{-0.021} $  &   $ -2.13\,^{+0.04}_{-0.05} $  &      $ 2.863\,^{+0.015}_{-0.015} $  &      $ 54.084\,^{+0.011}_{-0.011} $  &   3289.9  &   361  &   \\ \hline
      \hline\hline
    \end{tabular}
  \end{center}
\end{table*}
\begin{table*}
  \begin{center}
    \contcaption{}
    \renewcommand{\arraystretch}{2}
    \tiny
    \begin{tabular}{rlccccccccccc}
      \hline
      \hline
      {\bf ID}                        &
      {\bf GRB}                       &
      \multicolumn{3}{c}{{\bf Time sel. [s]}}  &
      {\bf Model}                     &
      {\bf Alpha}                     &
      {\bf Beta}                      &
      {\bf $\log E_{\rm p}$ [keV]}     &
      {\bf $\log E_{\rm iso}$ [erg]}   &
      {\bf C--STAT}                   &
      {\bf DOF}                       &\\
      \hline
 \multirow{2}{*}{32}  &   \multirow{2}{*}{100728B}  &     -1.2  &    -0.268  &       (first)  &   Band  &      $ -0.84\,^{+0.35}_{-0.29} $  &                      -2.3$^a$  &      $ 2.74\,^{+0.24}_{-0.15} $  &      $ 51.516\,^{+0.124}_{-0.092} $  &   592.61  &   484  &          \\
                      &                             &   -1.024  &     7.168  &       (whole)  &   Band  &      $ -0.85\,^{+0.12}_{-0.12} $  &      $ -2.3\,^{+0.2}_{-0.4} $  &   $ 2.547\,^{+0.070}_{-0.052} $  &      $ 52.610\,^{+0.050}_{-0.068} $  &   561.66  &   483  &   \\ \hline
 \multirow{2}{*}{33}  &   \multirow{2}{*}{100814A}  &     -0.6  &     0.132  &       (first)  &    CPL  &      $ -0.58\,^{+0.37}_{-0.27} $  &                                &      $ 3.09\,^{+0.24}_{-0.19} $  &         $ 51.48\,^{+0.17}_{-0.14} $  &   393.97  &   359  &          \\
                      &                             &   -1.152  &    152.45  &       (whole)  &    CPL  &   $ -0.536\,^{+0.100}_{-0.095} $  &                                &   $ 2.516\,^{+0.029}_{-0.027} $  &      $ 52.831\,^{+0.020}_{-0.019} $  &   737.63  &   359  &   \\ \hline
 \multirow{2}{*}{34}  &   \multirow{2}{*}{100906A}  &      0.2  &     1.018  &       (first)  &    CPL  &      $ -1.03\,^{+0.14}_{-0.21} $  &                                &      $ 3.25\,^{+0.62}_{-0.17} $  &         $ 52.12\,^{+0.24}_{-0.10} $  &      262  &   238  &          \\
                      &                             &        0  &    120.83  &       (whole)  &   Band  &   $ -1.307\,^{+0.078}_{-0.068} $  &   $ -2.00\,^{+0.08}_{-0.18} $  &                       $> 2.44 $  &                          $> 53.45 $  &   864.42  &   237  &   \\ \hline
 \multirow{1}{*}{35}  &   \multirow{1}{*}{101219A}  &           &            &   (short D14)  &         &      $ -0.22\,^{+0.27}_{-0.27} $  &                                &   $ 2.925\,^{+0.073}_{-0.088} $  &      $ 51.688\,^{+0.057}_{-0.065} $  &           &     0  &   \\ \hline
 \multirow{2}{*}{36}  &   \multirow{2}{*}{110106B}  &       -1  &    -0.515  &       (first)  &   Band  &      $ -0.83\,^{+0.77}_{-0.43} $  &                      -2.3$^a$  &                       $> 2.31 $  &                          $> 50.56 $  &   541.31  &   480  &          \\
                      &                             &   -2.048  &    18.432  &       (whole)  &    CPL  &      $ -1.11\,^{+0.13}_{-0.12} $  &                                &   $ 2.284\,^{+0.061}_{-0.052} $  &      $ 51.438\,^{+0.034}_{-0.032} $  &   538.47  &   480  &   \\ \hline
 \multirow{2}{*}{37}  &   \multirow{2}{*}{110128A}  &     -2.2  &    -1.198  &       (first)  &    CPL  &      $ -0.61\,^{+0.66}_{-0.46} $  &                                &      $ 2.80\,^{+0.29}_{-0.18} $  &         $ 51.37\,^{+0.17}_{-0.13} $  &   506.37  &   484  &          \\
                      &                             &   -2.048  &     9.216  &       (whole)  &    CPL  &      $ -1.01\,^{+0.30}_{-0.25} $  &                                &      $ 2.66\,^{+0.16}_{-0.12} $  &      $ 52.032\,^{+0.086}_{-0.075} $  &   477.39  &   484  &   \\ \hline
 \multirow{1}{*}{38}  &   \multirow{1}{*}{110213A}  &   -3.072  &     35.84  &       (whole)  &   Band  &      $ -1.29\,^{+0.12}_{-0.11} $  &   $ -2.13\,^{+0.06}_{-0.08} $  &   $ 2.142\,^{+0.070}_{-0.058} $  &      $ 52.983\,^{+0.024}_{-0.025} $  &   471.04  &   360  &   \\ \hline
 \multirow{2}{*}{39}  &   \multirow{2}{*}{110731A}  &     -0.8  &     0.349  &       (first)  &    CPL  &      $ -1.27\,^{+0.15}_{-0.14} $  &                                &   $ 2.665\,^{+0.068}_{-0.058} $  &      $ 52.355\,^{+0.040}_{-0.037} $  &   400.95  &   360  &          \\
                      &                             &   -1.024  &    13.312  &       (whole)  &   Band  &   $ -0.951\,^{+0.035}_{-0.034} $  &      $ -2.7\,^{+0.2}_{-0.3} $  &   $ 3.015\,^{+0.022}_{-0.021} $  &      $ 53.777\,^{+0.015}_{-0.017} $  &    501.5  &   359  &   \\ \hline
 \multirow{2}{*}{40}  &   \multirow{2}{*}{110818A}  &     -0.6  &     0.708  &       (first)  &   Band  &      $ -0.79\,^{+2.76}_{-0.55} $  &                      -2.3$^a$  &                       $> 2.34 $  &                          $> 51.95 $  &   479.49  &   481  &          \\
                      &                             &   -7.168  &    45.056  &       (whole)  &    CPL  &   $ -1.208\,^{+0.089}_{-0.084} $  &                                &   $ 3.109\,^{+0.113}_{-0.088} $  &      $ 53.292\,^{+0.052}_{-0.044} $  &   714.43  &   481  &   \\ \hline
 \multirow{1}{*}{41}  &   \multirow{1}{*}{111117A}  &           &            &   (short D14)  &         &      $ -0.28\,^{+0.28}_{-0.28} $  &                                &      $ 2.98\,^{+0.12}_{-0.18} $  &         $ 51.53\,^{+0.12}_{-0.16} $  &           &     0  &   \\ \hline
 \multirow{2}{*}{42}  &   \multirow{2}{*}{120119A}  &     -0.2  &     0.618  &       (first)  &    CPL  &      $ -1.30\,^{+0.92}_{-0.25} $  &                                &      $ 2.97\,^{+0.62}_{-0.53} $  &         $ 51.27\,^{+0.23}_{-0.28} $  &   393.34  &   360  &          \\
                      &                             &   -2.048  &    57.344  &       (whole)  &   Band  &   $ -0.941\,^{+0.027}_{-0.025} $  &      $ -2.5\,^{+0.1}_{-0.2} $  &   $ 2.710\,^{+0.018}_{-0.019} $  &      $ 53.568\,^{+0.018}_{-0.018} $  &   834.93  &   359  &   \\ \hline
 \multirow{2}{*}{43}  &   \multirow{2}{*}{120326A}  &       -1  &    -0.161  &       (first)  &    CPL  &      $ -1.05\,^{+0.27}_{-0.24} $  &                                &   $ 2.379\,^{+0.102}_{-0.080} $  &      $ 51.301\,^{+0.062}_{-0.057} $  &   514.08  &   483  &          \\
                      &                             &   -2.048  &    13.312  &       (whole)  &   Band  &      $ -0.87\,^{+0.19}_{-0.15} $  &      $ -2.5\,^{+0.1}_{-0.2} $  &   $ 2.118\,^{+0.038}_{-0.043} $  &      $ 52.546\,^{+0.027}_{-0.027} $  &   568.22  &   482  &   \\ \hline
 \multirow{3}{*}{44}  &   \multirow{3}{*}{120711A}  &      0.2  &     0.921  &       (prec.)  &    CPL  &      $ -0.35\,^{+0.52}_{-0.38} $  &                                &      $ 3.02\,^{+0.14}_{-0.11} $  &      $ 51.528\,^{+0.101}_{-0.093} $  &   368.65  &   362  &          \\
                      &                             &       61  &    61.722  &       (first)  &    CPL  &      $ -0.93\,^{+0.63}_{-0.19} $  &                                &      $ 3.45\,^{+0.25}_{-0.53} $  &         $ 51.72\,^{+0.14}_{-0.34} $  &   398.14  &   362  &          \\
                      &                             &   -1.024  &    117.76  &       (whole)  &   Band  &   $ -0.972\,^{+0.011}_{-0.011} $  &      $ -3.1\,^{+0.2}_{-0.2} $  &   $ 3.474\,^{+0.018}_{-0.017} $  &   $ 54.2786\,^{+0.0079}_{-0.0079} $  &    666.6  &   361  &   \\ \hline
 \multirow{2}{*}{45}  &   \multirow{2}{*}{120712A}  &     -1.4  &     0.152  &       (first)  &   Band  &      $ -1.25\,^{+0.45}_{-0.28} $  &                      -2.3$^a$  &                       $> 2.97 $  &                          $> 52.27 $  &   420.97  &   362  &          \\
                      &                             &     -1.4  &        16  &       (whole)  &   Band  &      $ -0.62\,^{+0.20}_{-0.20} $  &   $ -1.88\,^{+0.09}_{-0.14} $  &                       $> 2.81 $  &                          $> 53.30 $  &    468.5  &   361  &   \\ \hline
 \multirow{3}{*}{46}  &   \multirow{3}{*}{120716A}  &     -0.8  &     0.246  &       (prec.)  &   Band  &      $ -0.58\,^{+0.21}_{-0.19} $  &                      -2.3$^a$  &   $ 2.900\,^{+0.081}_{-0.068} $  &      $ 52.043\,^{+0.045}_{-0.042} $  &   376.43  &   361  &          \\
                      &                             &      177  &   178.046  &       (first)  &   Band  &      $ -0.58\,^{+0.23}_{-0.20} $  &                      -2.3$^a$  &   $ 3.008\,^{+0.103}_{-0.086} $  &      $ 52.092\,^{+0.049}_{-0.046} $  &   403.41  &   361  &          \\
                      &                             &      177  &       235  &       (whole)  &   Band  &   $ -1.077\,^{+0.061}_{-0.058} $  &      $ -2.7\,^{+0.2}_{-0.8} $  &   $ 2.660\,^{+0.036}_{-0.034} $  &      $ 53.325\,^{+0.030}_{-0.037} $  &   461.22  &   360  &   \\ \hline
 \multirow{2}{*}{47}  &   \multirow{2}{*}{120909A}  &     -1.6  &    -0.121  &       (first)  &    CPL  &      $ -0.76\,^{+0.24}_{-0.19} $  &                                &      $ 3.68\,^{+0.26}_{-0.22} $  &         $ 52.46\,^{+0.11}_{-0.14} $  &   376.75  &   360  &          \\
                      &                             &     -1.6  &       130  &       (whole)  &   Band  &   $ -1.039\,^{+0.063}_{-0.059} $  &      $ -2.2\,^{+0.1}_{-0.2} $  &   $ 2.941\,^{+0.049}_{-0.045} $  &      $ 53.818\,^{+0.029}_{-0.031} $  &   1086.8  &   359  &   \\ \hline
 \multirow{2}{*}{48}  &   \multirow{2}{*}{121128A}  &        2  &      2.96  &       (first)  &   Band  &      $ -0.43\,^{+0.21}_{-0.19} $  &                      -2.3$^a$  &   $ 2.596\,^{+0.054}_{-0.047} $  &      $ 52.099\,^{+0.028}_{-0.027} $  &   366.13  &   360  &          \\
                      &                             &     -0.6  &        28  &       (whole)  &   Band  &   $ -0.892\,^{+0.116}_{-0.095} $  &      $ -2.6\,^{+0.1}_{-0.2} $  &   $ 2.315\,^{+0.028}_{-0.033} $  &      $ 53.123\,^{+0.024}_{-0.024} $  &   491.99  &   359  &   \\ \hline
 \multirow{1}{*}{49}  &   \multirow{1}{*}{130427A}  &        0  &     0.402  &       (first)  &   Band  &   $ -0.399\,^{+0.026}_{-0.025} $  &      $ -3.5\,^{+0.3}_{-0.4} $  &   $ 3.060\,^{+0.015}_{-0.015} $  &      $ 51.817\,^{+0.011}_{-0.012} $  &    405.5  &   359  &   \\ \hline
 \multirow{2}{*}{50}  &   \multirow{2}{*}{130518A}  &       13  &    14.046  &       (first)  &   Band  &      $ -0.74\,^{+0.25}_{-0.23} $  &                      -2.3$^a$  &   $ 2.854\,^{+0.118}_{-0.084} $  &      $ 51.972\,^{+0.051}_{-0.046} $  &   481.17  &   481  &          \\
                      &                             &        0  &        75  &       (whole)  &   Band  &   $ -0.925\,^{+0.015}_{-0.014} $  &   $ -2.35\,^{+0.06}_{-0.07} $  &   $ 3.167\,^{+0.015}_{-0.015} $  &   $ 54.2712\,^{+0.0055}_{-0.0056} $  &   1128.5  &   480  &   \\ \hline
 \multirow{1}{*}{51}  &   \multirow{1}{*}{130603B}  &           &            &   (short D14)  &         &      $ -0.73\,^{+0.15}_{-0.15} $  &                                &   $ 2.952\,^{+0.061}_{-0.071} $  &      $ 51.326\,^{+0.045}_{-0.050} $  &           &     0  &   \\ \hline
 \multirow{2}{*}{52}  &   \multirow{2}{*}{130610A}  &      1.8  &     2.728  &       (first)  &    CPL  &      $ -0.50\,^{+0.46}_{-0.36} $  &                                &      $ 2.78\,^{+0.14}_{-0.10} $  &      $ 51.541\,^{+0.090}_{-0.080} $  &   340.54  &   360  &          \\
                      &                             &      1.8  &        20  &       (whole)  &    CPL  &      $ -0.83\,^{+0.11}_{-0.11} $  &                                &   $ 2.716\,^{+0.046}_{-0.042} $  &      $ 52.619\,^{+0.029}_{-0.027} $  &   467.98  &   360  &   \\ \hline
 \multirow{1}{*}{53}  &   \multirow{1}{*}{131004A}  &     -0.8  &       0.4  &       (short)  &   Band  &      $ -0.50\,^{+2.10}_{-0.63} $  &                      -2.3$^a$  &                       $> 1.71 $  &                          $> 51.13 $  &   368.42  &   360  &   \\ \hline
 \multirow{2}{*}{54}  &   \multirow{2}{*}{131011A}  &     -0.4  &     0.462  &       (first)  &    CPL  &      $ -0.71\,^{+0.30}_{-0.22} $  &                                &      $ 3.39\,^{+0.26}_{-0.21} $  &         $ 51.86\,^{+0.15}_{-0.15} $  &   475.15  &   480  &          \\
                      &                             &     -0.4  &        80  &       (whole)  &   Band  &   $ -0.997\,^{+0.168}_{-0.085} $  &      $ -2.0\,^{+0.2}_{-0.3} $  &                       $> 2.66 $  &                          $> 53.17 $  &   939.17  &   479  &   \\ \hline
 \multirow{2}{*}{55}  &   \multirow{2}{*}{131105A}  &     -0.4  &     0.406  &       (first)  &    CPL  &      $ -0.81\,^{+0.30}_{-0.24} $  &                                &      $ 3.03\,^{+0.23}_{-0.15} $  &         $ 51.47\,^{+0.14}_{-0.11} $  &   383.21  &   362  &          \\
                      &                             &     -0.4  &       120  &       (whole)  &    CPL  &   $ -1.295\,^{+0.029}_{-0.028} $  &                                &   $ 2.795\,^{+0.036}_{-0.033} $  &      $ 53.280\,^{+0.017}_{-0.016} $  &   704.34  &   362  &   \\ \hline
 \multirow{2}{*}{56}  &   \multirow{2}{*}{131108A}  &      0.2  &      1.22  &       (first)  &   Band  &   $ -0.549\,^{+0.101}_{-0.090} $  &      $ -2.1\,^{+0.1}_{-0.1} $  &   $ 2.933\,^{+0.049}_{-0.050} $  &      $ 52.872\,^{+0.022}_{-0.024} $  &    366.2  &   361  &          \\
                      &                             &      0.2  &        22  &       (whole)  &   Band  &   $ -0.867\,^{+0.030}_{-0.029} $  &      $ -2.4\,^{+0.1}_{-0.2} $  &   $ 3.048\,^{+0.025}_{-0.024} $  &      $ 53.799\,^{+0.013}_{-0.014} $  &   458.83  &   361  &   \\ \hline
 \multirow{2}{*}{57}  &   \multirow{2}{*}{131231A}  &      1.8  &     2.293  &       (first)  &    CPL  &   $ -0.046\,^{+0.839}_{-0.580} $  &                                &   $ 2.418\,^{+0.128}_{-0.091} $  &      $ 50.256\,^{+0.093}_{-0.086} $  &    403.2  &   360  &          \\
                      &                             &      1.8  &        70  &       (whole)  &   Band  &   $ -1.208\,^{+0.011}_{-0.010} $  &   $ -2.33\,^{+0.04}_{-0.05} $  &   $ 2.453\,^{+0.011}_{-0.011} $  &   $ 53.3352\,^{+0.0082}_{-0.0086} $  &   1261.2  &   359  &          \\
      \hline\hline
    \end{tabular}
  \end{center}
\end{table*}

\section{Comparsion with Gruber et al. (2014) results}
\label{cmp-g14}

\begin{figure*}
  \includegraphics[width=.6\textwidth]{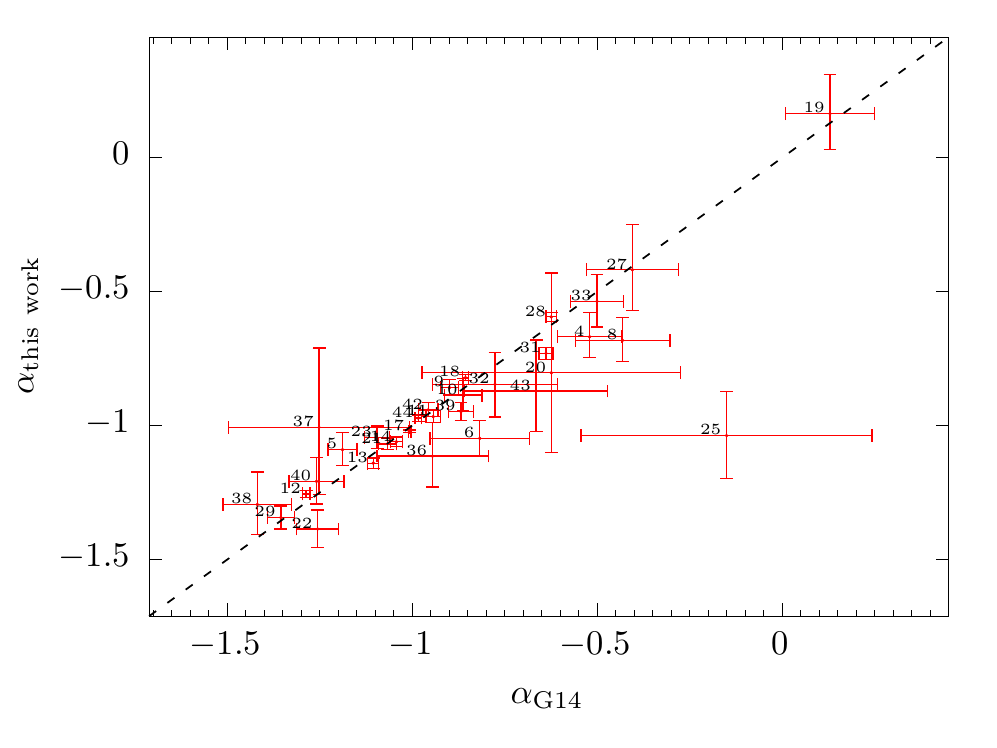}\\
  \vspace{-0.5cm}  
  \includegraphics[width=.6\textwidth]{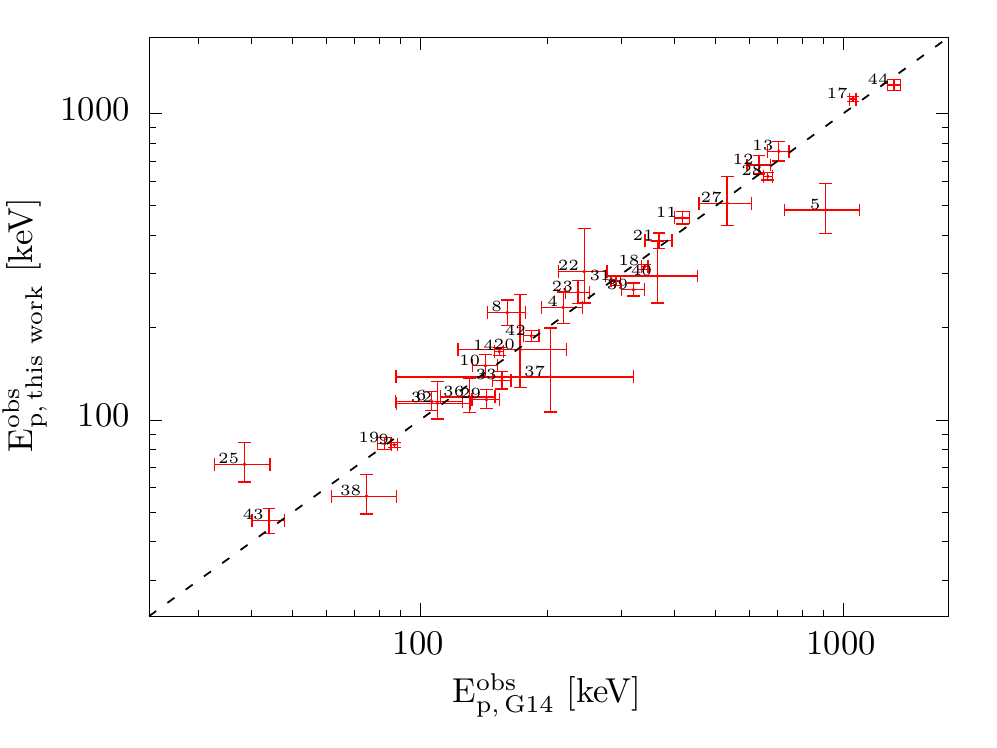}\\
  \vspace{-0.5cm}
  \includegraphics[width=.6\textwidth]{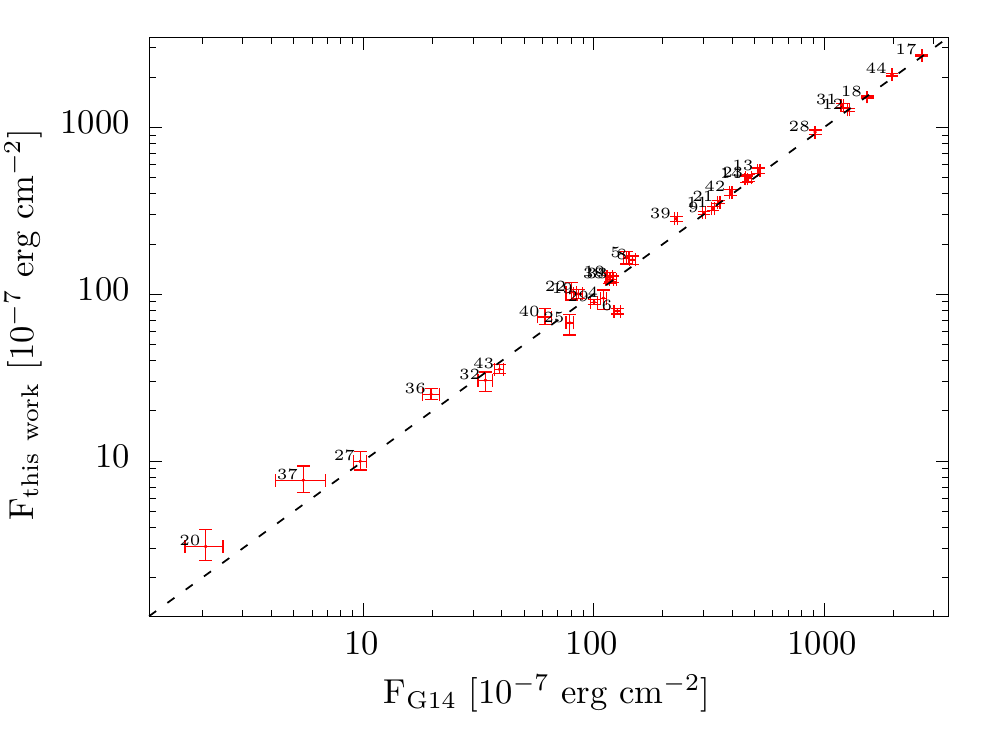}
  \caption{Comparison of the results of our {\it short} and {\it
      whole} analysis to those of \citet{2014-Gruber-GBMCat-4yr}, for
    the bursts present in both samples.  Upper panel: low energy
    spectral index \texttt{alpha}.  Middle panel: observed $\nu
    F_{\nu}$ peak energy.  Lower panel: fluence estimated in the 10
    keV--1 MeV (observer frame) energy range.  The dashed line are the
    1:1 lines.}
  \label{fig-cmpg14}
\end{figure*}

In this section we compare the results of our {\it short} and {\it
  whole} analysis to those of \citet{2014-Gruber-GBMCat-4yr}, for the
bursts present in both samples.  In Fig. \ref{fig-cmpg14} we compare
the values of the low energy spectral index \texttt{alpha} (upper
panel), the observed $\nu F_{\nu}$ peak energy (middle panel) and the
fluence estimated in the 10 keV--1 MeV (observer frame) energy range
(lower panel).

The large discrepancies found in the \texttt{alpha} parameter for GRB
080810 (ID 5) and in the observed peak energy GRB 091208B (ID 25) are
likely related to the very high background contamination.  The
discrepancy in fluence for GRB 080916A (ID 6) is due to the different
value of the \texttt{beta} parameter used in
\citet{2014-Gruber-GBMCat-4yr}.  Also this burst is significantly
background dominated.

\label{lastpage}
\end{document}